\documentclass[12pt]{article}

\usepackage[latin1]{inputenc}
\usepackage{amsmath}
\usepackage{graphicx}
\usepackage{cite}
\usepackage{color}

\graphicspath{ {figs/} }

\allowdisplaybreaks

\begin{document}

\title{\vskip-3cm{\baselineskip14pt
    \begin{flushright}
      \normalsize DESY 16-018, HU-EP-16/02, TTP16-004
  \end{flushright}}
  \vskip1.0cm 
  Electron contribution to the muon anomalous magnetic moment at four loops}

\author{
  Alexander~Kurz$^{a,b}$,
  Tao~Liu$^c$,
  Peter~Marquard$^b$,
  Alexander V.~Smirnov$^d$,\\
  Vladimir A.~Smirnov$^{e,f}$,
  Matthias Steinhauser$^a$
  \\[1em]
  {\small\it $(a)$ Institut f{\"u}r Theoretische Teilchenphysik, 
    Karlsruhe Institute of Technology (KIT),}\\
  {\small\it 76128 Karlsruhe, Germany}
  \\
  {\small\it $(b)$ Deutsches Elektronen Synchrotron DESY, Platanenallee 6}\\
  {\small\it 15738 Zeuthen, Germany}
  \\
  {\small\it $(c)$ Department of Physics, University of Alberta,}\\
  {\small\it Edmonton AB T6G 2J1, Canada}
  \\
  {\small\it $(d)$ Scientific Research Computing Center, Moscow State University,}\\
  {\small\it 119991, Moscow, Russia}
  \\
  {\small\it $(e)$ Skobeltsyn Institute of Nuclear Physics of Moscow State University,}\\
  {\small\it 119991, Moscow, Russia}
   \\
  {\small\it $(f)$ Institut f\"{u}r Mathematik und Institut f\"{u}r Physik,}\\ 
  {\small\it Humboldt-Universit\"{a}t zu Berlin, 12489 Berlin, Germany}
}

\date{}

\maketitle

\begin{abstract}
  We present results for the QED contributions to the anomalous
  magnetic moment of the muon containing closed electron loops. The
  main focus is on perturbative corrections at four-loop order where
  the external photon couples to the external muon. Furthermore, all
  four-loop contributions involving simultaneously a closed electron
  and tau loop are computed. In combination with our recent
  results on the light-by-light-type corrections (see
  Ref.~\cite{Kurz:2015bia}) the complete four-loop electron-loop
  contribution to the anomalous magnetic moment of the muon has been
  obtained with an independent calculation.
  Our calculation is based on an asymptotic expansion in the ratio of
  the electron and the muon mass and shows the importance of higher
  order terms in this ratio. We perform a detailed comparison with
  results available in the literature and find good numerical
  agreement.
  As a by-product we present analytic results for the on-shell muon
  mass and wave function renormalization constants at three-loop order
  including massive closed electron and tau loops, which we also calculated
  using the method of asymptotic expansion.
\end{abstract}


\thispagestyle{empty}

\pagebreak


\section{Introduction}

The anomalous magnetic moment of the muon, $a_\mu$, is an important
observable in particle physics. Both on the
experimental~\cite{Bennett:2006fi,Roberts:2010cj} and the theory side
a lot of effort has been invested to provide precise results for
$a_\mu$. The theory prediction can be split into hadronic, electroweak
and QED contribution. The non-perturbative hadronic contribution is
further subdivided into the vacuum
polarization~\cite{Davier:2010nc,Hagiwara:2011af,Jegerlehner:2011ti,Benayoun:2012wc,Krause:1996rf,Greynat:2012ww,Hagiwara:2003da,Kurz:2014wya}
and light-by-light
contribution~\cite{Nyffeler:2009tw,Melnikov:2003xd,Bijnens:2007pz,Colangelo:2014qya} 
and has reached next-to-next-to-leading order accuracy. It is
nevertheless the main source to the uncertainty of the theory
prediction.  On the other hand, the electroweak part is known up to
two-loop
order~\cite{Czarnecki:1995sz,Knecht:2002hr,Czarnecki:2002nt,Gnendiger:2013pva}
and thus well under control. The numerically largest contribution
arises from QED corrections. Up to three loops analytical results are
available~\cite{Schwinger:1948iu,Petermann:1957hs,Sommerfield:1957zz,Elend:1966,Samuel:1990qf,Laporta:1992pa,Laporta:1993ju,Laporta:1996mq,Czarnecki:1998rc,Passera:2006gc}
and four and five-loop corrections have been computed in
Refs.~\cite{Kinoshita:2004wi,Aoyama:2007mn,Aoyama:2012wk} using
numerical methods.  In this paper we complete the cross-check of the
four-loop corrections involving at least one closed electron loop
using an independent method.

We cast the perturbative expansion of $a_\mu$ in the form
\begin{eqnarray}
  a_\mu &=& \sum_{n=1}^\infty a_\mu^{(2n)} \left( \frac{\alpha}{\pi} \right)^n
  \,,
  \label{eq::amu}
\end{eqnarray}
where $n$ counts the number of loops. It is common practice to further split
the four-loop term into four parts according to
\begin{eqnarray}
  a_\mu^{(8)} &=& A_1^{(8)} + A_2^{(8)}(m_\mu / m_e) + A_2^{(8)}(m_\mu / m_\tau) 
  + A_3^{(8)}(m_\mu / m_e, m_\mu / m_\tau)
  \,,
  \label{eq::Amu}
\end{eqnarray}
where $A_1^{(8)}$ denotes the universal part which includes the pure
photonic corrections and closed muon loops.  The arguments of the remaining
three contributions indicate which leptons are involved in the corresponding
Feynman diagrams.  Note that analytic results for $A_2^{(8)}(m_\mu / m_\tau)$
have been computed in Ref.~\cite{Kurz:2013exa} and the light-by-light
contributions of $A_2^{(8)}(m_\mu / m_e)$ are presented in
Ref.~\cite{Kurz:2015bia}.  In this work we will concentrate on the
non-light-by-light contributions of $A_2^{(8)}(m_\mu / m_e)$. Furthermore, we
present analytic results for $A_3^{(8)}(m_\mu / m_e, m_\mu / m_\tau)$.  For
convenience we split the last three contributions in Eq.~(\ref{eq::Amu}) into
a so-called light-by-light part and a remainder and write
\begin{eqnarray}
  A_{2/3}^{(8)} &=& A_{2/3}^{(8),\rm lbl} + A_{2/3}^{(8),\rm rem}
  \,.
\end{eqnarray}

\begin{figure}[tb]
  \begin{center}
    \begin{tabular}{ccc}
      \includegraphics[scale=0.45]{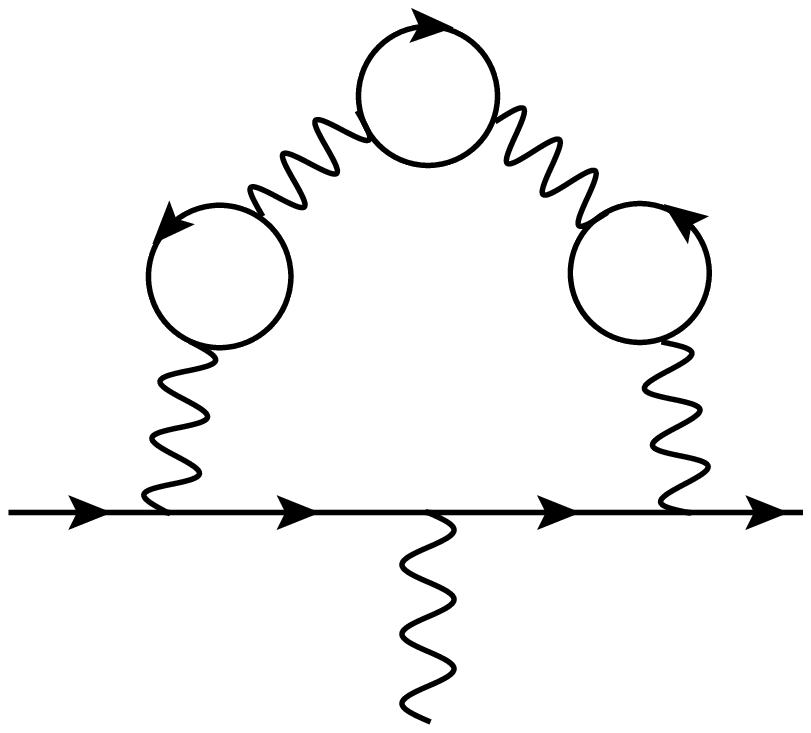}
      & \includegraphics[scale=0.45]{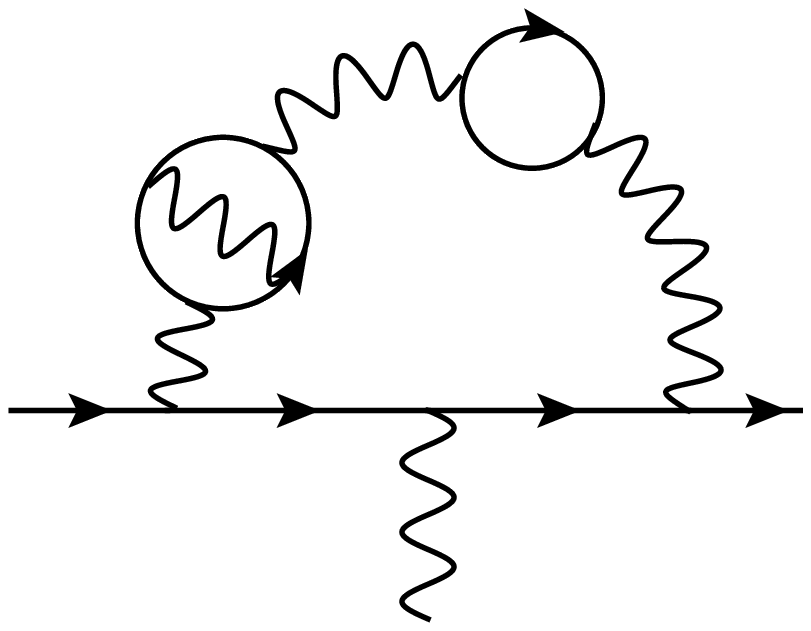}
      & \includegraphics[scale=0.45]{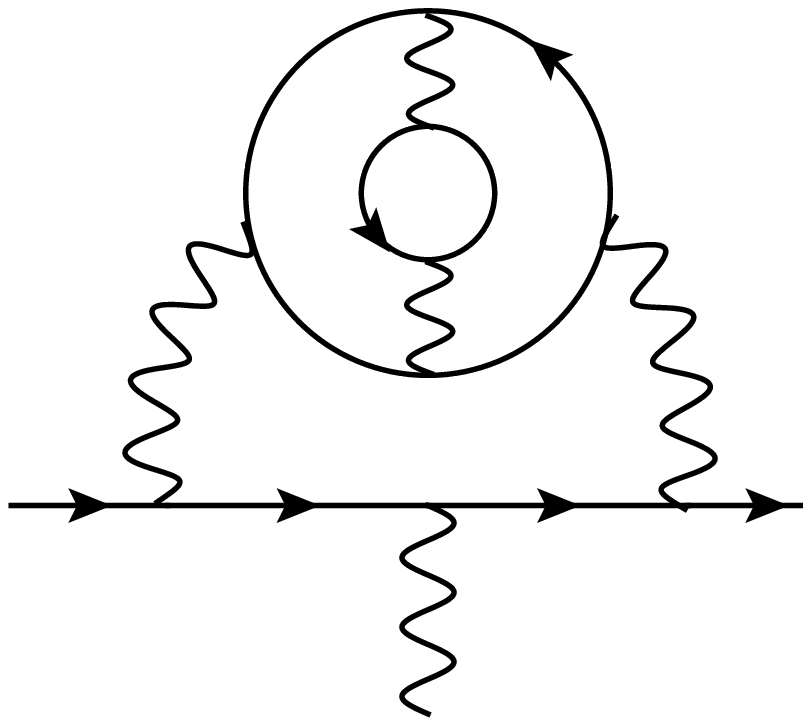} \\ 
      I(a) & I(b) & I(c) \\
      \includegraphics[scale=0.45]{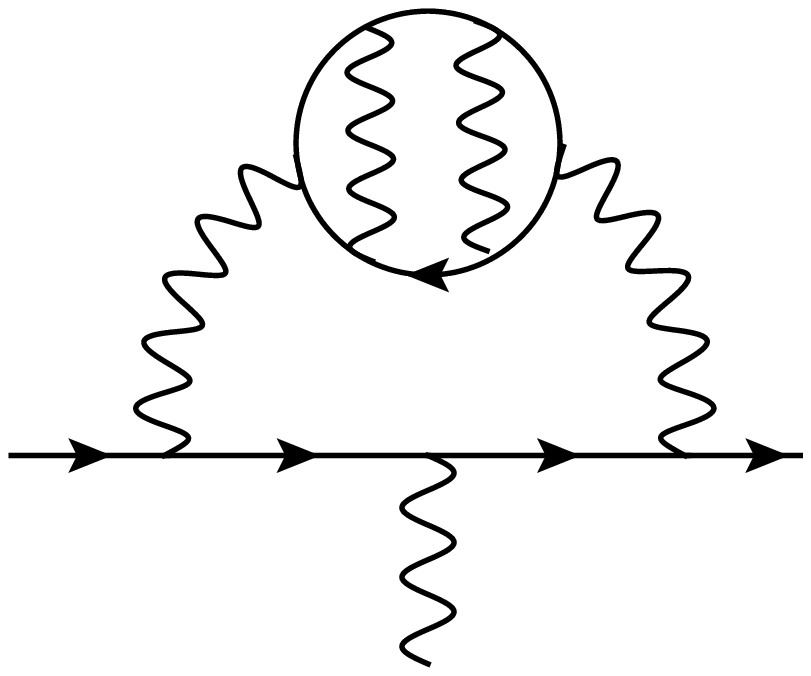}
      & \includegraphics[scale=0.45]{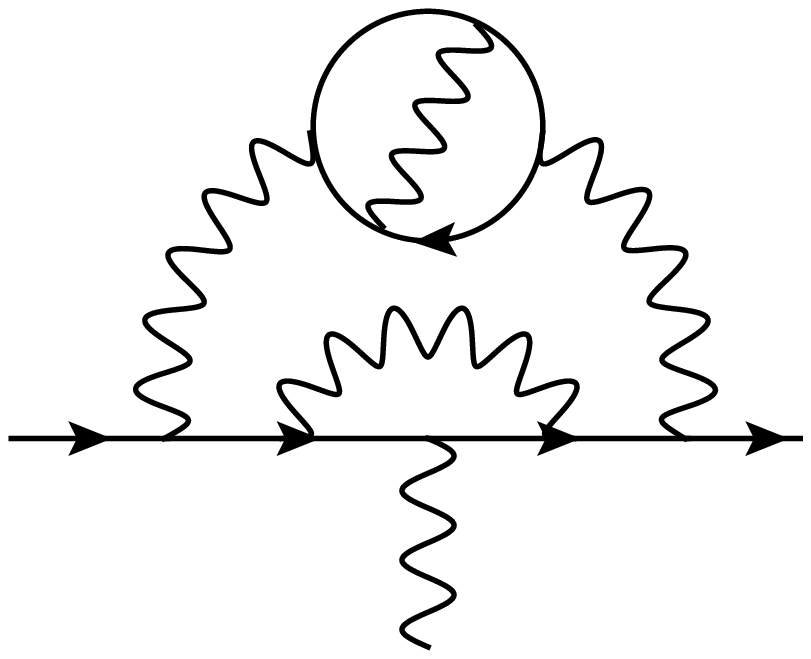}
      & \includegraphics[scale=0.45]{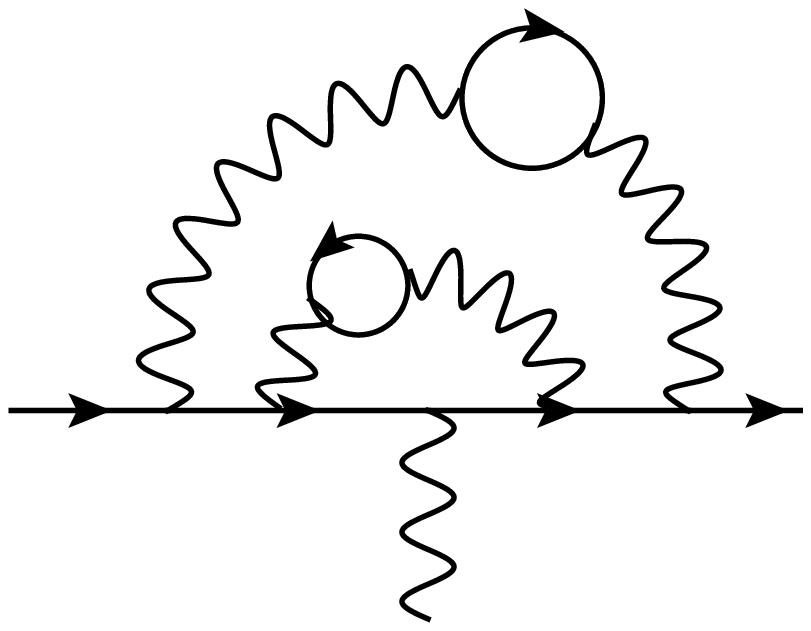} \\ 
      I(d) & II(a) & II(b) \\
      \includegraphics[scale=0.45]{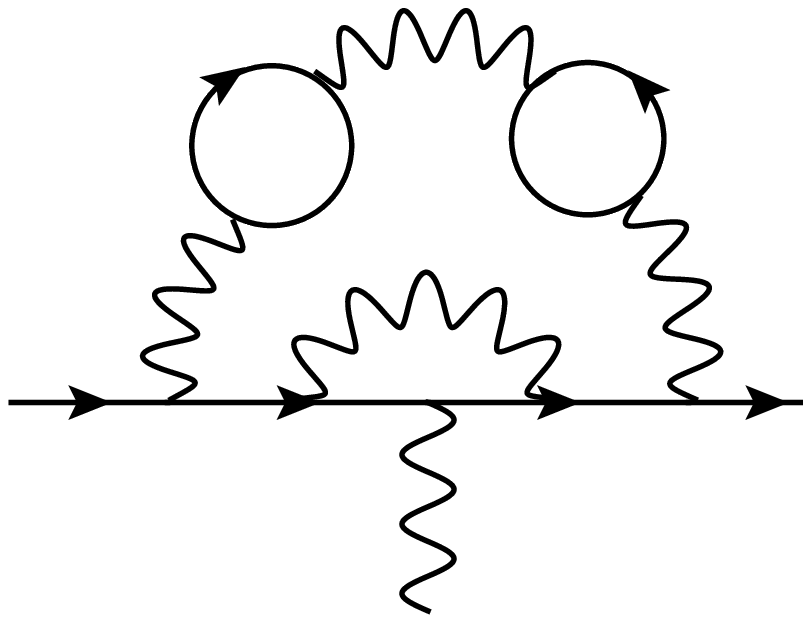}
      & \includegraphics[scale=0.45]{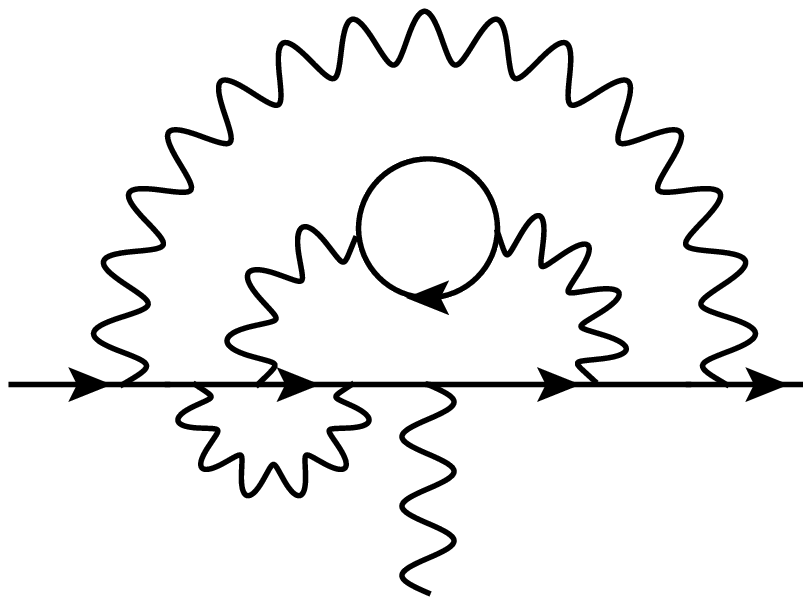}
      & \includegraphics[scale=0.45]{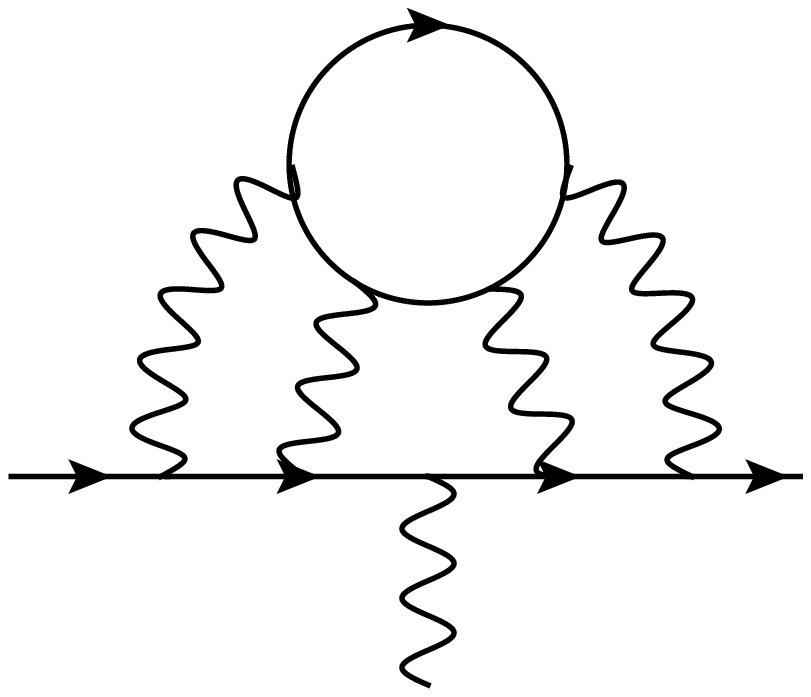} \\ 
      II(c) & III & IV(d)
      \\
      \includegraphics[scale=0.45]{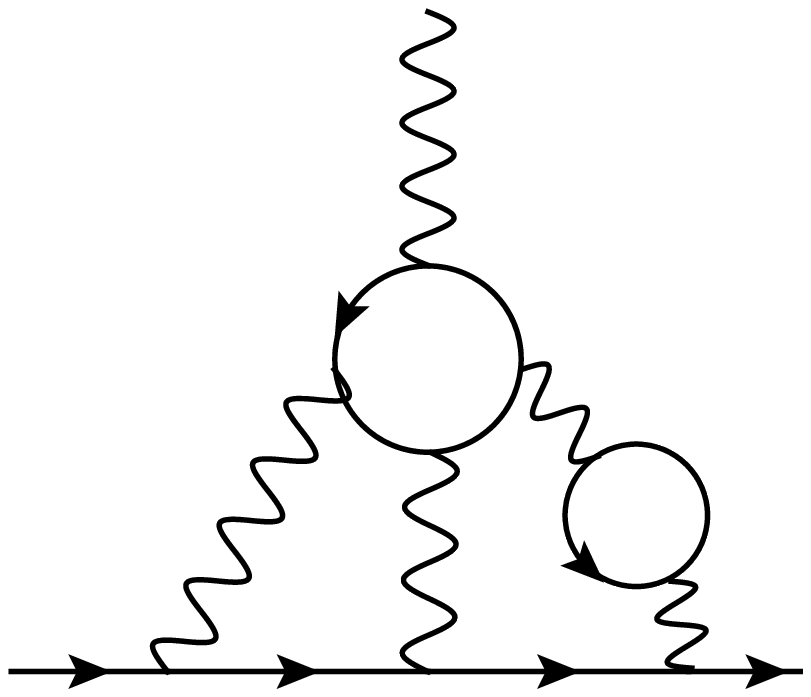}
      & \includegraphics[scale=0.45]{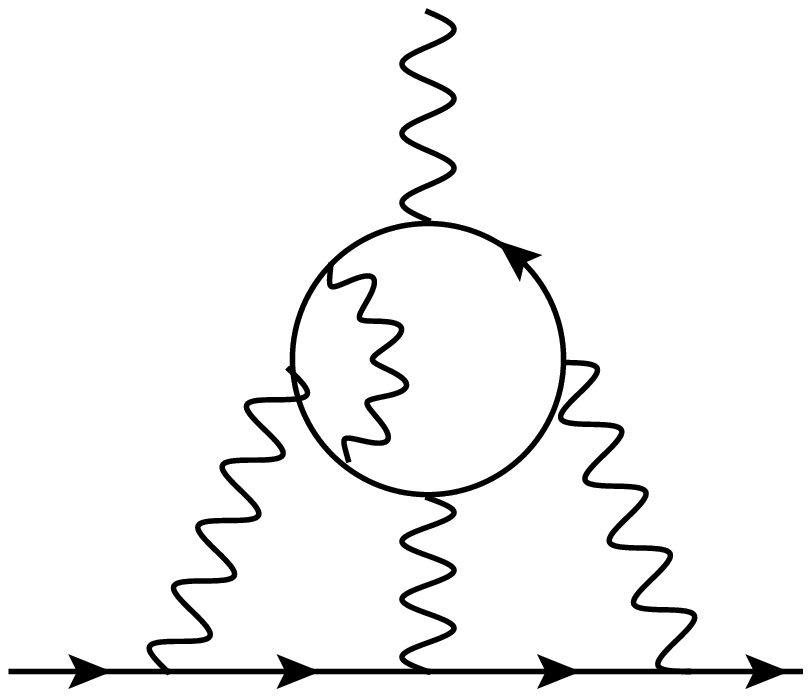}
      & \includegraphics[scale=0.45]{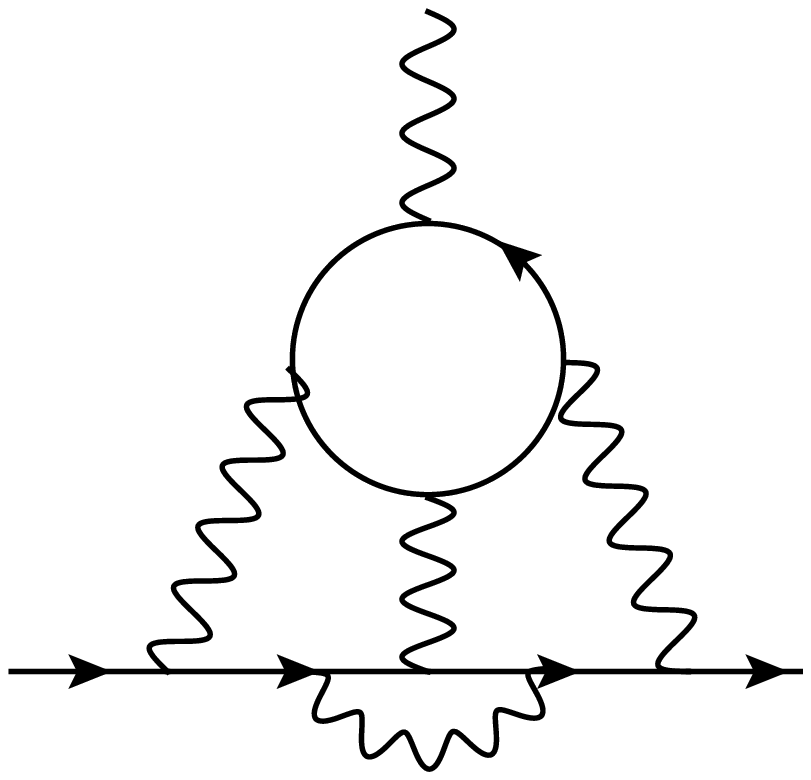}
      \\
      IV(a) & IV(b) & IV(c)
    \end{tabular}
  \end{center}
  \caption{
    Four-loop example Feynman diagrams contributing to $a_\mu$ 
    containing at least one closed electron loop. The
    external solid lines represent muons, the solid loops denote
    electrons, muons or taus, and the wavy lines represent photons.}
  \label{fig::classes}
\end{figure}

In Figure~\ref{fig::classes} we define (following Ref.~\cite{Aoyama:2012wk}) a
subdivision into different classes of Feynman diagrams which are individually
finite and gauge invariant. All of them contribute to $A_2^{(8)}(m_\mu /
m_e)$, those with two closed fermion loops also to $A_3^{(8)}(m_\mu / m_e,
m_\mu / m_\tau)$.  The light-by-light-type cases are denoted by IV(a), IV(b)
and IV(c).

In contrast to $A_2^{(8),\rm lbl}(m_\mu / m_e)$ it is possible to
obtain the leading term for $m_e/m_\mu\to0$ of $A_2^{(8),\rm
  rem}(m_\mu / m_e)$ by simply setting the electron mass to zero. One
obtains a finite result after renormalizing the coupling constant
$\alpha$ in the $\overline{\rm MS}$ scheme and thus gets an expression
which depends on $\log(\mu/m_\mu)$.  Here, $\mu$ is the
renormalization scale which also appear in the argument of
$\bar{\alpha}(\mu)$.  The transformation of $\bar\alpha$ to the
on-shell scheme introduces $\log(\mu/m_e)$ terms in such a way that
the renormalization scale $\mu$ drops out and a dependence on
$\log(m_\mu/m_e)$ remains.  This approach has been used to get
analytic results for the leading term to $A_2^{(8),\rm
  rem}(m_\mu/m_e)$ involving two or three electron loops which
constitute contributions to I(a), I(b), I(c), II(b) and II(c).  In
this paper we complement these results with higher order terms in the
$m_e/m_\mu$ expansion and add numerical results for all remaining
contributions of $A_2^{(8),\rm rem}(m_\mu/m_e)$. The corresponding
results are presented in Section~\ref{sec::res}.

In Section~\ref{sec::res_e_tau} we discuss the contribution
$A_3^{(8)}(m_\mu / m_e, m_\mu / m_\tau)$, which is computed by
applying a nested asymptotic expansion for the hierarchy $m_e\ll m_\mu
\ll m_\tau$. Analytic results are presented and compared to the
numerical results present in the literature.

The structure of the paper is as follows: In the next section we
discuss some of the calculational challenges we met during our
calculation. Afterwards we discuss in Section~\ref{sec::ren} the
renormalization procedure and present new results for the three-loop
renormalization constants of the muon mass and wave function which are
not yet available in the literature.  After presenting results for
$a_\mu$ in Sections~\ref{sec::res} and~\ref{sec::res_e_tau} we
conclude in Section~\ref{sec::concl}.  We dedicate the appendix to
useful analytic results for some of the four-loop coefficients of
$a_\mu$ (Appendix~\ref{app::A}).


\section{\label{sec::calc}Calculation details}

The integrals which have to be computed for the quantities considered
in this paper are on-shell integrals where the square of the external
momentum equals $m_\mu^2$. As a further scale one has the electron mass
$m_e$ and for the corrections of Section~\ref{sec::res_e_tau} also the
tau mass $m_\tau$ is present. Since there is a strong hierarchy among
the lepton masses, $m_e\ll m_\mu\ll m_\tau$ we apply an asymptotic
expansion to rewrite the two-scale integrals in terms of products of
one-scale integrals. This procedure provides the renormalization
constants and the anomalous magnetic moment in terms of a series
expansion in the mass ratios. The asymptotic expansion significantly
simplifies the complexity of the underlying integrals and sometimes
even leads to analytic results.

We apply the strategy of expansion by regions
\cite{Beneke:1997zp,Smirnov:2002pj,Pak:2010pt,Jantzen:2012mw} (see also
a recent review in Chapter~9 of \cite{Smirnov:2012gma}) which provides
an asymptotic expansion of a given Feynman integral in a given limit  
represented as a finite sum of contributions corresponding to so-called regions
(i.e. scalings of components of loop momenta or Feynman parameters).
Each term of such contributions is manifestly homogeneous with respect
to the expansion parameter.

\begin{figure}[tb]
  \begin{center}
    \begin{tabular}{lc}
      full diagram: & \raisebox{-0.8cm}{\includegraphics[scale=0.5]{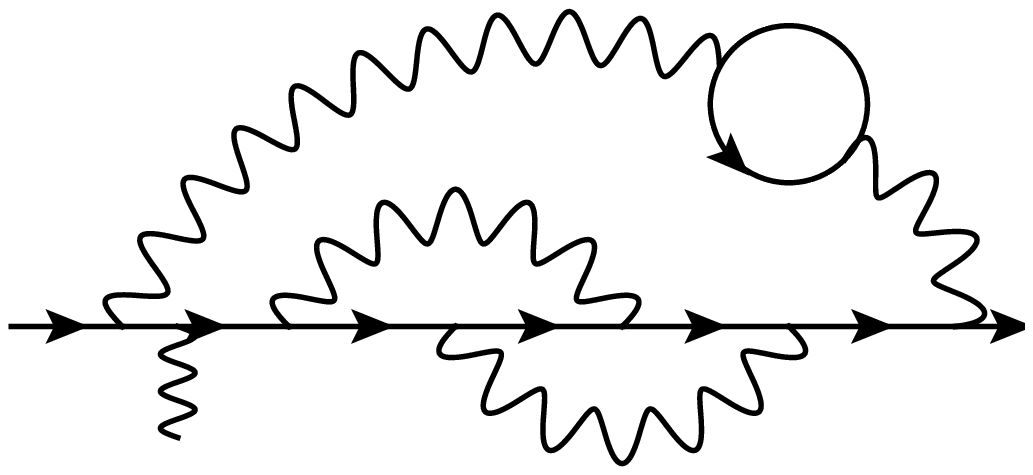}} \\
      1. all-hard region: & \raisebox{-0.8cm}{\includegraphics[scale=0.5]{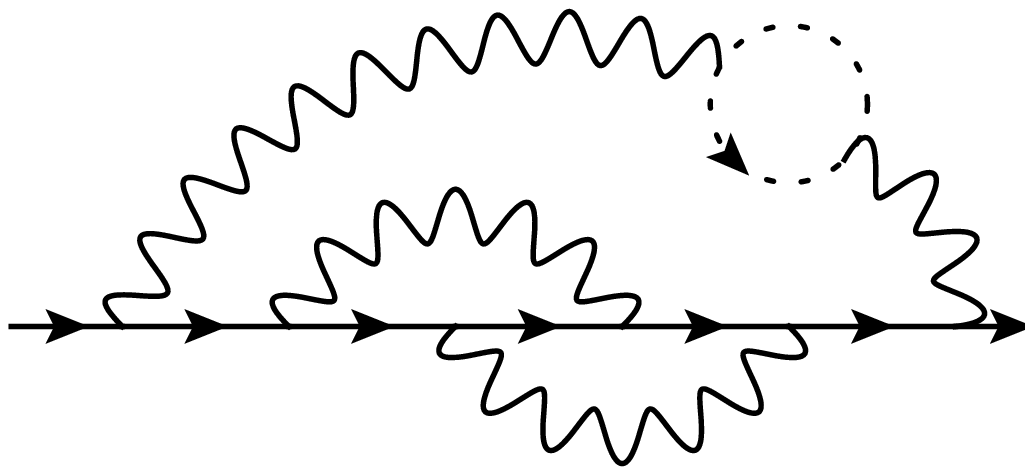}} \\
      2. three hard momenta: & \raisebox{-0.8cm}{\includegraphics[scale=0.5]{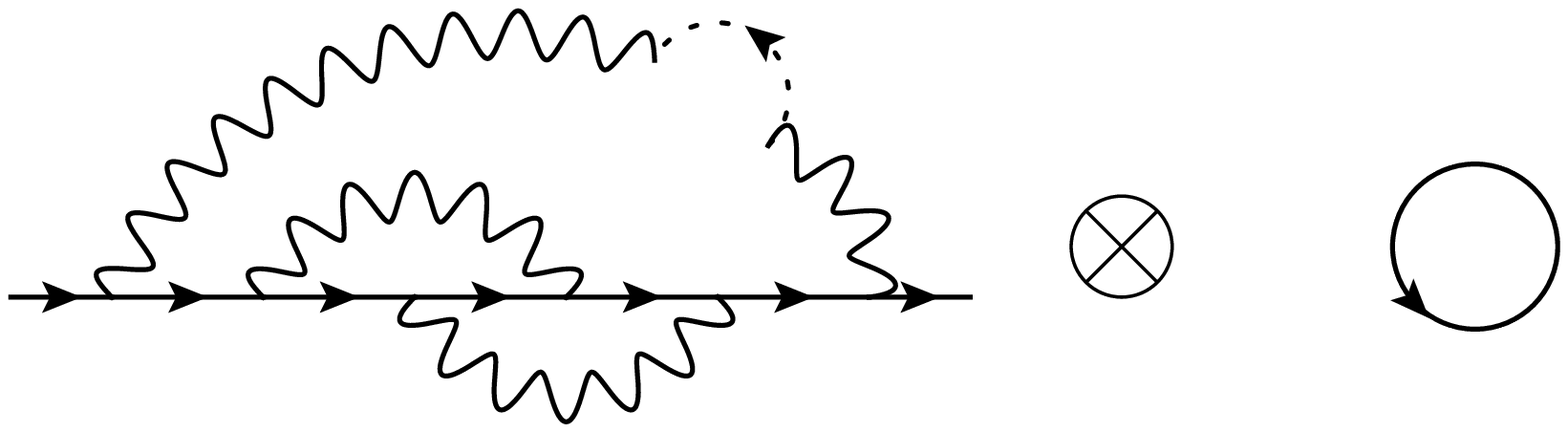}} \\
      3. two hard momenta: & \includegraphics[scale=0.5]{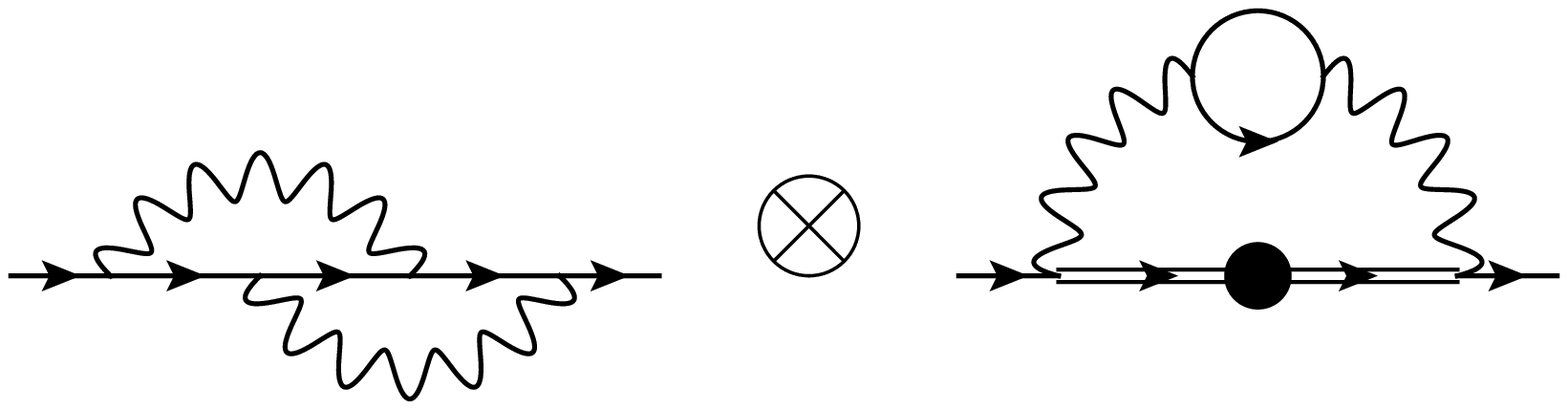} \\
      4. one hard momentum: & \raisebox{-0.8cm}{\includegraphics[scale=0.5]{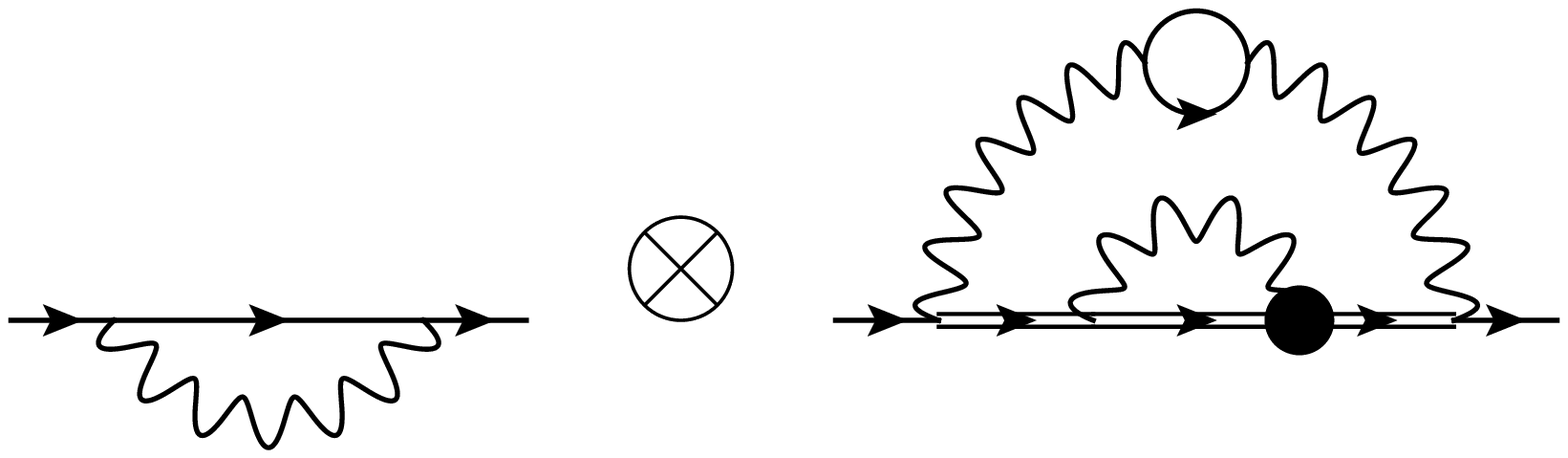}} \\
      5. all-soft region: & \raisebox{-0.8cm}{\includegraphics[scale=0.5]{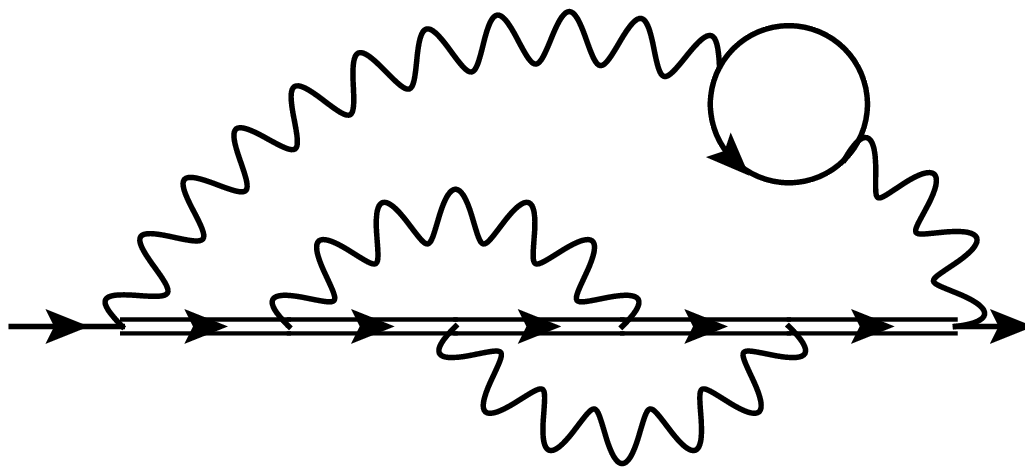}}
    \end{tabular}
  \end{center}
  \caption{Relevant regions of an example Feynman diagram with an electron
    loop. The first line contains the full diagram and in rows two to six one
    representative sub-diagram for each region is shown.  The
    dotted lines denote massless (electron) propagators, the solid lines
    massive (muon or electron) propagators and the solid double lines
    represent linear propagators of the form $1/(2 \ell \cdot q)$ where $q$ is
    the external momentum with $q^2=m_\mu^2$ and $\ell$ is a loop momentum.}
  \label{fig::asy}
\end{figure}

In Figure~\ref{fig::asy} we demonstrate the way we apply the asymptotic
expansion to a typical diagram. The sub-diagrams shown in rows two to
six are obtained by allowing each loop momentum to scale either as
the hard scale, i.e., $m_\mu$, or as the soft scale which is given by
the electron mass. This leads to regions which determine
the expansion prescription for each propagator.
For the diagram in Figure~\ref{fig::asy} one
obtains five types of contributions which we discuss in the following:
\begin{enumerate}

\item In case all loop momenta are hard one obtains a four-loop 
  on-shell integral. As the electron propagators contain
  heavy momenta, a Taylor expansion in $m_e$ has to be performed and
  thus the electron propagators become massless.

\item There are two contributions corresponding to regions where one of the electron propagators is
  soft and the momenta through all other propagators are hard.  This
  situation is obtained in case the loop momentum in the electron loop
  is of order $\ell_1 \sim m_e$ and none of the other momenta are flowing through
  one of the electron propagators. This propagator leads to a one-loop vacuum
  integral with mass $m_e$. The other part of the diagram becomes a
  three-loop on-shell integral. Note that the second electron
  propagator has to be expanded in $\ell_1$ and $m_e$
  which looks as follows
  \begin{eqnarray}
    \frac{1}{(\ell_1 + \ell_2)^2 - m_e^2} &=& \frac{1}{\ell_2^2}
    \sum_{n=0}^\infty \left( \frac{-\ell_1^2 - 2\ \ell_1\cdot\ell_2 +
      m_e^2}{\ell_2^2} \right)^n
    \,.
  \end{eqnarray} 

\item A further contribution corresponds to the region, where the two loop momenta through
  the photon lines which are not connected to the electron loop are
  hard. This leads to a two-loop on-shell integral. In the remaining
  two-loop integral where both loop momenta scale like $m_e$ 
  one has to choose the external momentum along the muon lines.
  The corresponding propagators have to be expanded 
  according to ($\ell_1\sim m_e$)
  \begin{eqnarray}
    \frac{1}{(\ell_1 + q)^2 - m_\mu^2} &=& \frac{1}{\ell_1^2 + 2\ell_1
      \cdot q} = \frac{1}{2\ell_1 \cdot q} \sum_{n=0}^\infty
    \left( \frac{-\ell_1^2}{2\ell_1 \cdot q} \right)^n
    \,.
  \end{eqnarray}
  The linear propagator $1/(2\ell_1 \cdot q)^n$ does not introduce
  an additional scale, because it only gives rise to an overall factor
  $(q^2)^{-n/2}$. The rest of the integral is independent of
  $q^2$ and also $m_\mu$.

\item If one replaces in the previous region one of the hard loop
  momenta by a soft one, one obtains a one-loop on-shell integral and 
  a three-loop linear integral.

\item In the last region all loop momenta are of order $m_e$ and the
  external muon momentum must again flow through the muon line. This
  leads to a four-loop integral which has only the scale $m_e$;
  $m_\mu$ only appears as a trivial pre-factor.

\end{enumerate}

The relevant regions can be found by examining the scaling behaviour
of the alpha-parameter representation of a given Feynman
diagram~\cite{Smirnov:2002pj}. This approach is implemented in the
Mathematica package {\tt asy}~\cite{Pak:2010pt,Jantzen:2012mw}.  An
alternative approach, which is implemented in an in-house Mathematica
program, is based on the fact that each loop momentum is either hard
or soft.  More details are given in Ref.~\cite{Kurz:2015bia} where the
light-by-light-type four-loop contribution has been computed.

Let us mention that a non-trivial issue is the calculation of
the tensor integrals
which occur due to the factorization of the integrals. The most
complicated cases which we had to implement for our calculation 
were tensor integrals up to rank ten for three-loop vacuum integrals 
and up to rank ten and eight for one- and two-loop on-shell integrals,
respectively.

The reduction of the tensor structure as well as the evaluation of
traces of $\gamma$-matrices is done with {\tt
  FORM}~\cite{Vermaseren:2000nd,Kuipers:2012rf} and {\tt
  TFORM}~\cite{Tentyukov:2007mu} (see also
Ref.~\cite{Steinhauser:2015wqa}).  The scalar integrals are reduced
with {\tt FIRE}~\cite{Smirnov:2014hma} and {\tt
  crusher}~\cite{crusher} to a relative small set of
master integrals.\footnote{The four-loop on-shell master integrals are
  a subset of the ones which are needed for the quark mass relation
  considered in Ref.~\cite{Marquard:2015qpa}. In addition there are
  128 four-loop ``linear masters'' with at least one 
  propagator of the form $1/(2 \ell \cdot q)$.} 
Some of them could be evaluated analytically or to
high numerical precision. The $\epsilon$ expansion of the remaining
ones is computed numerically with the help of {\tt
  FIESTA}~\cite{Smirnov:2013eza}. Further details can be found in
Ref.~\cite{Kurz:2015bia}.
Let us stress that our approach leads to an analytic expression for
$a_\mu$ as a linear combination of master integrals. Thus, if
necessary, the accuracy of the final result can be systematically
improved.


\section{On-shell renormalization constants}\label{sec::ren}

In this section we discuss the renormalization procedure and provide
results for the three-loop renormalization constants which are not yet
available in the literature.

The four-loop calculation performed in this paper requires the on-shell
renormalization of the muon mass and wave function to three-loop order.
We adopt the same notation as for $a_\mu$ and define
\begin{eqnarray}
  Z_{2/m}^{\rm OS} &=& Z_{2/m,1}^{\rm OS} 
  + \delta Z_{2/m,2}^{\rm OS}(m_\mu / m_e) 
  + \delta Z_{2/m,2}^{\rm OS}(m_\mu / m_\tau) 
  + \delta Z_{2/m,3}^{\rm OS}(m_\mu / m_e, m_\mu / m_\tau)
  \,.
  \label{eq::Z2m}
\end{eqnarray}
Up to two loops the mass and wave function renormalization constant can be
found in Refs.~\cite{Gray:1990yh,Broadhurst:1991fy}. At three-loop order only
the limit $m_e\to0$ exists in analytic
form~\cite{Melnikov:2000qh,Marquard:2007uj,Melnikov:2000zc} (see also
Refs.~\cite{Chetyrkin:1999ys,Chetyrkin:1999qi} where $Z_m^{\rm OS}$ has been
computed for the first time using numerical methods) whereas the fermionic
corrections are only known in numerical
form~\cite{Bekavac:2007tk}.\footnote{Analytic, asymptotically expanded
  three-loop results for the light-fermion contribution to $Z_2^{\rm OS}$ can
  be found in Ref.~\cite{Bekavac:2009gz}.} In this section we
present analytic results for $\delta Z_{2/m,2}^{\rm OS}(m_\mu / m_e)$
including correction terms in $m_e/m_\mu$. They are needed to construct the
result for $a_\mu$ shown in Section~\ref{sec::res}.  Furthermore, we also
present the results for $\delta Z_{2/m,3}^{\rm OS}(m_\mu / m_e, m_\mu /
m_\tau)$ which are needed for the construction of $a_\mu$ from
Section~\ref{sec::res_e_tau}.  The results for $\delta Z_{2/m,2}^{\rm
  OS}(m_\mu / m_\tau)$ can be found in Ref.~\cite{Kurz:2013exa}.

Note that the renormalization of the electron mass is needed
up to two loops. The corresponding counterterm can be obtained from
Refs.~\cite{Gray:1990yh,Bekavac:2007tk}.

As a further ingredient the on-shell renormalization constant of $\alpha$
is needed up to three loops. In the practical calculation we renormalize the
coupling in a first step in the $\overline{\rm MS}$ scheme
and transform afterwards to the on-shell scheme. 
The corresponding conversion formula with massive leptons can be
obtained from Ref.~\cite{Lee:2013sx}.

\begin{figure}[tb]
  \begin{center}
    \begin{tabular}{ccc}
      \includegraphics[scale=0.45]{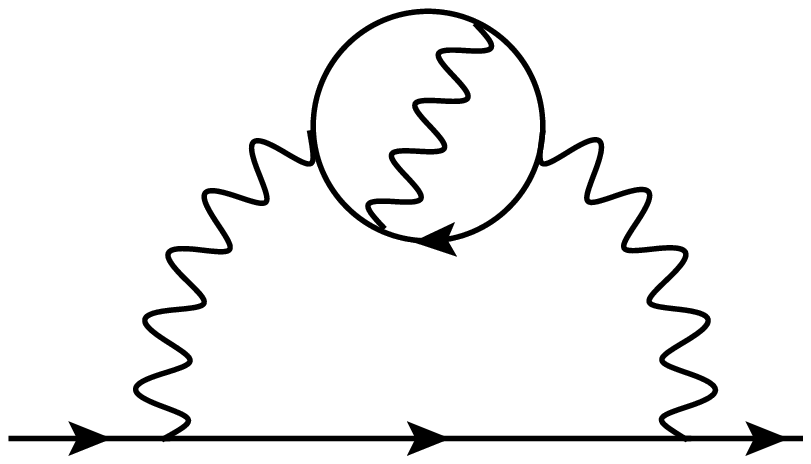} &
      \includegraphics[scale=0.45]{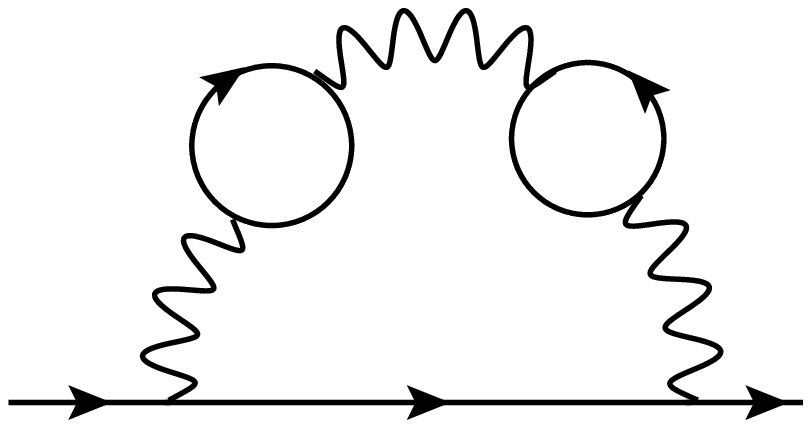} &
      \includegraphics[scale=0.45]{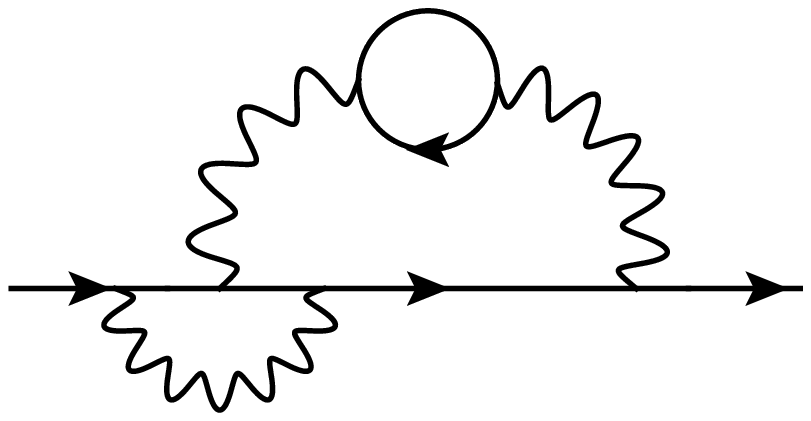}
    \end{tabular}
  \end{center}
  \caption{Propagator type diagrams with electron loops contributing to
    the mass and field renormalization constants. The middle diagram can
    contain two electron loops or one electron and one muon loop.}
  \label{fig::ren}
\end{figure}

Sample diagrams which contribute to $Z_m^{\rm OS}$ and $Z_2^{\rm OS}$
are shown in Figure~\ref{fig::ren}. For the computation of the corresponding
integrals we apply the asymptotic expansion described in the previous 
section and obtain the result as a series in $m_e/m_\mu$.

In the following we present the three-loop electron contributions to
$Z_m^{\rm OS}$ and $Z_2^{\rm OS}$ para\-metrized in terms of the fine
structure constant in the $\overline{\rm MS}$ scheme,
$\bar\alpha$. The masses $m_e$ and $m_\mu$ are renormalized on-shell.
The symbols $\ell_i$ with $i=e,\mu$ denote $\log ( \mu^2 / m_i^2 )$,
where $\mu$ is the renormalization scale. Moreover we label closed
electron and muon loops by $n_e$ and $n_\mu$,
respectively. Furthermore, $\zeta_j$ denotes Riemann's zeta function
and $a_4 = \text{Li}_4 (1/2)$.  Our results for three-loop
contributions to $\delta Z_{m,2}^{\rm OS}(m_\mu / m_e)$ and $\delta
Z_{2,2}^{\rm OS}(m_\mu / m_e)$ read
\begin{align*}
\delta^{(3)} Z_{m,2}^{\rm OS}
=\ & \left( \frac{\bar\alpha}{\pi} \right)^3 \bigg( n_e n_\mu \bigg[ -\frac{1}{18 \epsilon^3} +\frac{5}{108 \epsilon^2}+\frac{35}{648 \epsilon} -\frac{\ell_\mu^3}{18} -\frac{13 \ell_\mu^2}{36} +\frac{13 \pi^2}{108} \\
& +\left(\frac{\pi^2}{18}-\frac{143}{108}\right) \ell_\mu +\frac{2 \zeta_3}{9}-\frac{5917}{3888} +\frac{\pi^2 \ell_\mu m_e}{6 m_\mu} +\frac{m_e^2}{m_\mu^2} \left(-\ell_\mu-\frac{2}{3}\right) \\
& +\frac{m_e^3}{m_\mu^3} \left(\frac{\ell_\mu \pi^2}{6}+\frac{4 \pi^2}{45}\right) \bigg] + n_e \bigg[\frac{3}{32 \epsilon^3}+\frac{1}{\epsilon^2} \bigg( \frac{3 \ell_\mu}{32} -\frac{5}{192} \bigg) +\frac{1}{\epsilon} \bigg( -\frac{3 \ell_\mu^2}{64} \\
& -\frac{7 \pi^2}{128}-\frac{23 \ell_\mu}{64}-\frac{\zeta_3}{4}-\frac{65}{384}  +\frac{3 \pi^2 m_e}{16 m_\mu} -\frac{9 m_e^2}{8 m_\mu^2} +\frac{3 \pi^2 m_e^3}{16 m_\mu^3} \bigg) \\
& +\ell_\mu \left(-\frac{ \pi^2 \log (2)}{3}+\frac{5 \pi^2}{384}-\frac{\zeta_3}{4}-\frac{497}{384}\right) -\frac{11 \ell_\mu^3}{64} -\frac{117 \ell_\mu^2}{128} +\frac{\log^4 (2)}{9} \\
& +\frac{2 \pi^2 \log^2 (2)}{9}-\frac{11 \pi^2 \log (2)}{9}-\frac{119 \pi^4}{2160}+\frac{1091 \pi^2}{2304}+\frac{8 a_4}{3} +\frac{145 \zeta_3}{96} \\
& +\frac{575}{2304} +\frac{m_e}{m_\mu} \bigg( \frac{13 \pi^3}{18}+\frac{3 \ell_e \pi^2}{8}+\frac{3 \ell_\mu \pi^2}{16}+\frac{5 \pi^2 \log (2)}{18} -\frac{751 \pi^2}{432} \bigg) \\
& +\frac{m_e^2}{m_\mu^2} \bigg( \frac{3 \pi^2}{4}-\frac{9 \ell_e}{4}-\frac{9 \ell_\mu}{8}-\frac{51}{8} \bigg)+\frac{m_e^3}{m_\mu^3} \bigg(\frac{7 \pi^3}{12}+\frac{11 \ell_e \pi^2}{72} +\frac{59 \ell_\mu \pi^2}{144} \\
& +\frac{5 \pi^2 \log (2)}{18}-\frac{49 \pi^2}{36} \bigg) \bigg] +n_e^2 \bigg[ -\frac{1}{36 \epsilon^3}+\frac{5}{216 \epsilon^2}+\frac{35}{1296 \epsilon} -\frac{\ell_\mu^3}{36} \\
& -\frac{13 \ell_\mu^2}{72} +\left(-\frac{\pi^2}{18}-\frac{89}{216}\right) \ell_\mu -\frac{13 \pi^2}{108}-\frac{7 \zeta_3}{18}-\frac{2353}{7776} \\
& +\frac{m_e}{m_\mu} \left(\frac{\ell_e \pi^2}{6}+\frac{4 \pi^2}{45}\right)
+\frac{m_e^2}{m_\mu^2} \left(-\ell_\mu-\frac{2}{3}\right)+\frac{\ell_e \pi^2
  m_e^3}{6 m_\mu^3} \bigg] 
{+ {\cal O}\left(\frac{m_e^4}{m_\mu^4}\right)}
\bigg)\,,
\end{align*}
\begin{align*}
\delta^{(3)} Z_{2,2}^{\rm OS}
=\ & \left( \frac{\bar\alpha}{\pi} \right)^3 \bigg( n_e n_\mu \bigg[\frac{1}{36 \epsilon^2} +\frac{1}{\epsilon} \bigg( -\frac{\ell_e \ell_\mu}{6}-\frac{5}{216} \bigg) -\frac{\ell_\mu^3}{12}-\frac{19 \ell_\mu^2}{36}-\frac{\ell_e^2 \ell_\mu}{6} \\
& +\left(-\frac{355}{108}+\frac{11 \pi^2}{72}\right) \ell_\mu +\ell_e \left(-\frac{\ell_\mu^2}{12}+\frac{2 \ell_\mu}{9}-\frac{\pi^2}{72}\right)+\frac{13 \pi^2}{36}-\frac{4721}{1296} \\
& +\frac{\pi^2 \ell_\mu m_e}{4 m_\mu}+\frac{m_e^2}{m_\mu^2} \bigg(\frac{\ell_e}{5} -\frac{11 \ell_\mu}{5} -\frac{13}{25}\bigg) +\frac{m_e^3}{m_\mu^3} \left(\frac{5 \pi^2 \ell_\mu}{12}+\frac{2 \pi^2}{15}\right) \bigg] \\
& + n_e \bigg[ \frac{1}{\epsilon^2} \bigg( -\frac{3 \ell_e}{16}-\frac{7}{192} \bigg) + \frac{1}{\epsilon} \bigg( -\frac{3 \ell_e^2}{16}-\frac{3 \ell_\mu^2}{32}+\ell_e \left(\frac{3}{16}-\frac{3 \ell_\mu}{16}\right) \\
& -\frac{41 \ell_\mu}{64}-\frac{5 \pi^2}{64}-\frac{203}{384} +\frac{9 \pi^2 m_e}{32 m_\mu} -\frac{9 m_e^2}{4 m_\mu^2}+\frac{15 \pi^2 m_e^3}{32 m_\mu^3} \bigg) -\frac{\ell_e^3}{8} -\frac{3 \ell_\mu^3}{16} \\
& +\left(\frac{9}{32}-\frac{3 \ell_\mu}{16}\right) \ell_e^2 -\frac{179 \ell_\mu^2}{128} +\left(-\frac{3 \ell_\mu^2}{32}+\frac{3 \zeta_3}{4}-\frac{3 \pi^2}{64}+\frac{29}{48}\right) \ell_e +\frac{28 a_4}{3} \\
& +\frac{641 \zeta_3}{96}+\frac{7 \log^4(2)}{18} +\ell_\mu \left(\frac{\zeta_3}{4}-\frac{2}{3} \pi^2 \log (2)+\frac{31 \pi^2}{96}-\frac{537}{128}\right) \\
& +\frac{5}{18} \pi^2 \log^2(2)-\frac{47}{18} \pi^2 \log (2) -\frac{59 \pi^4}{432}+\frac{2881 \pi^2}{2304}-\frac{5875}{2304} \\
& +\frac{m_e}{m_\mu} \bigg(\frac{9 \pi^2 \ell_e}{16}+\frac{5}{12} \pi^2 \log (2)+\frac{13 \pi^3}{12} +\frac{9 \ell_\mu \pi^2}{32}-\frac{751 \pi^2}{288}\bigg) \\
& +\frac{m_e^2}{m_\mu^2} \bigg(\left(\frac{\ell_\mu}{2}-\frac{\ell_e^2}{4}-\frac{67}{12} \right) \ell_e-\frac{\ell_\mu^2}{4}-\frac{7 \ell_\mu}{6}+\frac{17 \pi^2}{8} -\frac{1267}{72}\bigg) \\
& +\frac{m_e^3}{m_\mu^3} \bigg(\frac{47 \pi^2 \ell_e}{144}+\frac{41}{36} \pi^2 \log (2)+\frac{35 \pi^3}{24}+\frac{311 \ell_\mu \pi^2}{288} -\frac{47 \pi^2}{12}\bigg) \bigg] \\
& + n_e^2 \bigg[\frac{1}{72 \epsilon^2}+\frac{1}{\epsilon} \bigg(-\frac{\ell_e^2}{12}-\frac{5}{432}\bigg)-\frac{5 \ell_e^3}{36}+\frac{\ell_e^2}{9}-\frac{\ell_\mu^3}{36}-\frac{19 \ell_\mu^2}{72} \\
& +\left(-\frac{31}{108}-\frac{\pi^2}{72}\right) \ell_e+\ell_\mu \left(-\frac{167}{216}-\frac{\pi^2}{18}\right)-\frac{19 \pi^2}{108}-\frac{2449}{2592} \\
& +\frac{m_e}{m_\mu} \left(\frac{\pi^2 \ell_e}{4}+\frac{2
    \pi^2}{15}\right)+\frac{m_e^2}{m_\mu^2} \left(-2
  \ell_\mu-\frac{7}{3}\right)+\frac{5 \pi^2 \ell_e m_e^3}{12 m_\mu^3} \bigg] 
{+ {\cal O}\left(\frac{m_e^4}{m_\mu^4}\right)}
\bigg)\,.
\end{align*} 
$\delta^{(3)} Z_{m,2}^{\rm OS}$ and $\delta^{(3)} Z_{2,2}^{\rm OS}$ agree with
the numerical results given in Ref.~\cite{Bekavac:2007tk} and $\delta^{(3)}
Z_{2,2}^{\rm OS}$ agrees with the analytic expression of
Ref.~\cite{Bekavac:2009gz}.

For $\delta Z_{m,3}^{\rm OS}(m_\mu / m_e, m_\mu / m_\tau)$ and $\delta
Z_{2,3}^{\rm OS}(m_\mu / m_e, m_\mu / m_\tau)$ we obtain
\begin{align*}
\delta Z_{m,3}^{{\rm OS}} =& 
{\left( \frac{\bar\alpha}{\pi} \right)^3}
n_e n_\tau \bigg[\text{$-$}\frac{1}{18 \epsilon^3}+\frac{5}{108 \epsilon^2}+\frac{35}{648 \epsilon}+\frac{\ell_\tau^3}{36} +\ell_\tau \left(-\frac{1}{8}-\frac{13 \ell_\mu}{36}-\frac{\ell_\mu^2}{12}-\frac{\pi^2}{18}\right) \\
&-\frac{1327}{3888}+\frac{2 \zeta_3}{9}+\frac{\ell_\tau \pi^2}{6}\frac{m_e}{m_\mu}-\frac{\ell_\tau m_e^2}{m_\mu^2}+\left(-\frac{46}{75}-\frac{\ell_\mu}{5}+\frac{\ell_\tau}{5}\right)\frac{m_e^2}{m_\tau^2} \\
&+\left(\frac{529}{3375}-\frac{2 \ell_\mu}{45}-\frac{\ell_\mu^2}{45}-\frac{\ell_\tau}{25}+\frac{\ell_\tau^2}{45}-\frac{2 \pi^2}{135}\right)\frac{m_\mu^2}{m_\tau^2} \\
&+\frac{\ell_\tau \pi^2}{6}\frac{m_e^3}{m_\mu^3}+\frac{4 \pi^2 m_e^3}{45 m_\mu m_\tau^2} +\left(-\frac{142}{3675}-\frac{\ell_\mu}{35}+\frac{\ell_\tau}{35}\right)\frac{m_e^2 m_\mu^2}{m_\tau^4} \\
&+\left(\frac{552889}{16464000}-\frac{9 \ell_\mu}{1120}-\frac{3 \ell_\mu^2}{560}-\frac{37 \ell_\tau}{9800}+\frac{3 \ell_\tau^2}{560}-\frac{\pi^2}{280}\right)\frac{m_\mu^4}{m_\tau^4} \\
&+\left(\frac{99227}{7501410}-\frac{38 \ell_\mu}{14175}-\frac{2 \ell_\mu^2}{945}-\frac{199 \ell_\tau}{297675}+\frac{2 \ell_\tau^2}{945}-\frac{4 \pi^2}{2835}\right)\frac{m_\mu^6}{m_\tau^6} \\
&+\left(-\frac{286}{33075}-\frac{\ell_\mu}{105}+\frac{\ell_\tau}{105}\right)\frac{m_e^2
  m_\mu^4}{m_\tau^6}
+\ldots\bigg]
\,,
\end{align*}
\begin{align*}
\delta Z_{2,3}^{{\rm OS}} =& 
{\left( \frac{\bar\alpha}{\pi} \right)^3}
n_e n_\tau \bigg[\frac{1}{36 \epsilon^2}-\frac{5}{216 \epsilon}-\frac{\ell_e \ell_\tau}{6\epsilon}-\frac{\ell_e^2 \ell_\tau}{6}-\frac{35}{1296} +\ell_e \left(\frac{2 \ell_\tau}{9}-\frac{\ell_\tau^2}{12}-\frac{\pi^2}{72}\right) \\
&+\ell_\tau \left(-\frac{229}{216}-\frac{19 \ell_\mu}{36}-\frac{\ell_\mu^2}{12}-\frac{5 \pi^2}{72}\right) +\frac{\ell_\tau \pi^2}{4}\frac{m_e}{m_\mu} -\frac{2 \ell_\tau m_e^2}{m_\mu^2}\\
&+\left(-\frac{2}{45}-\frac{2 \ell_\mu}{45}\right)\frac{m_\mu^2}{m_\tau^2} +\left(-\frac{1}{2}+\frac{\ell_e}{5}-\frac{\ell_\mu}{5} \right)\frac{m_e^2}{m_\tau^2} +\frac{5 \ell_\tau \pi^2}{12}\frac{m_e^3}{m_\mu^3} \\
&+\frac{2 \pi^2}{15}\frac{m_e^3}{m_\mu m_\tau^2} -\frac{m_e^2 m_\mu^2}{35\,m_\tau^4} +\bigg(-\frac{685189}{16464000}-\frac{3 \ell_\mu}{1120}+\frac{3 \ell_\mu^2}{560} \\
&+\frac{37 \ell_\tau}{9800}-\frac{3 \ell_\tau^2}{560}+\frac{\pi^2}{280}\bigg)\frac{m_\mu^4}{m_\tau^4} +\left(-\frac{29}{33075}+\frac{\ell_\mu}{105}-\frac{\ell_\tau}{105}\right)\frac{m_e^2 m_\mu^4}{m_\tau^6} \\
&+\left(\frac{16 \ell_\mu}{14175}+\frac{4 \ell_\mu^2}{945}+\frac{398
    \ell_\tau}{297675}-\frac{4 \ell_\tau^2}{945}+\frac{8
    \pi^2}{2835}-\frac{546409}{18753525}\right)\frac{m_\mu^6}{m_\tau^6} 
+\ldots\bigg]
\,,
\end{align*}
where the ellipses stand for higher orders in $1/m_\tau$ and $m_e$.


\section{Electron contribution to $a_\mu$}\label{sec::res}

In this section we present results for the four-loop contribution
$A_2^{(8)}(m_\mu / m_e)$ which involves one or more closed electron
loops. The application of asymptotic expansion leads to a series
expansion in the parameter $x = m_e/m_\mu = 1/206.76\ldots$. Since $x\ll1$
the convergence of the expansion is in general quite
fast as we demonstrate by evaluating the first four expansion terms.
In general it is necessary to include the linear term in $x$ to obtain
a result with sub-percent accuracy. On the other hand, the term of
order $x^3$ are negligible in all cases.

In the following we discuss in detail all diagram classes
(cf. Figure~\ref{fig::classes}) contributing to $A_{2}^{(8),\rm
  rem}(m_\mu / m_e)$; the light-by-light-type results can be found in
Ref.~\cite{Kurz:2015bia}.

We illustrate for each class how the numerical values of the final
results are build up from the expansion terms in $x  
\approx 1/206.7682843(52)$ and
the logarithmic contributions. First the expansion in $x$ and its
dependence on $\ell_x = \log (x)$ is shown. After the first equality
sign the numerical values of $x$ and $\ell_x \approx -5.33\ldots$ are
inserted, but the resulting summands are kept separated, which
indicates the relative behaviour between the constant and the
logarithmic terms. Afterwards the sums for every order in $x$ are
evaluated, so the convergence of the asymptotic series is
demonstrated. At the end the final contribution of the diagram class
is shown.

In the course of the numerical evaluation of $a_\mu$ there are 
several sources for numerical uncertainties which originate from
numerical Monte-Carlo integrations. We interpret them as statistical
one-sigma uncertainties and combine them in quadrature to arrive at
the final uncertainty.

Diagram class I(a) consists of diagrams with three fermion loop
insertions in one photon line. According to the type of the internal
fermions we split this class further into I(a0), I(a1) and I(a2) where
in the case of I(a0) all fermion loops are electrons, the subclass
I(a1) has one muon loop and two electron loops, and in the case of
I(a2) two of the loops consist of muons and just one is an electron
loop. The results for the three sub-classes read
\begin{align*}
A_2^\text{(8),I(a0)} =\ & -3.266377 - 2.687872 \ell_x - 0.925926 \ell_x^2 - 0.148148 \ell_x^3 \\
 & + x[6.40516] \\
 & + x^2[17.24475 + 20.03224 \ell_x + 5.77778 \ell_x^2 + 1.77778 \ell_x^3] \\
 & + x^3[-52.0022] \\
=\ & [-3.26638 + 14.33065 - 26.32032 + 22.45270] \\
 & + [0.0309775] \\
 & + [0.000403 - 0.002498 + 0.003842 - 0.006302] \\
 & + [-0.000005883] \\
=\ & [7.19666] + [0.03098] + [-0.004555] + [-0.000005883] \\
=\ & 7.22308
\,,
\\
A_2^\text{(8),I(a1)} =\ & 0.0167998 + 0.0220046 \ell_x + 0.0209166 \ell_x^2 \\
 & + x[0] \\
 & + x^2[-0.548361 - 0.254463 \ell_x] \\
 & + x^3[0] \\
=\ & [0.016700 - 0.117320 + 0.594573] + [0] \\
 & + [-0.00001283 + 0.00003173] + [0] \\
=\ & [0.494053] + [0] + [0.00001891] + [0] \\
=\ & 0.494072
\,, 
\\
A_2^\text{(8),I(a2)} =\ & 0.000706151 - 0.00511705 \ell_x \\
 & + x[0] \\
 & + x^2[0.00493387] \\
 & + x^3[0] \\
=\ & [0.0007062 + 0.0272821] + [0] \\
 & + [0.0000001154] + [0] \\
=\ & [0.0279882] + [0] + [0.0000001154] + [0] \\
=\ & 0.0279883
\,.
\end{align*}
Due to the three electron loops the $x^0$ term of I(a0) has a cubic
logarithmic correction. One observes sizeable cancellations among the
individual logarithmic contributions which results in a final
contribution of about 7.2.  Note that the odd powers in $x$ have no
logarithmic enhancement terms.  As a result, the higher order
corrections in $x$ lead to a shift at the sub-percent level. 
The contributions with one or two muon loops are numerically
less important. On the one hand this is connected to the lower
maximal logarithmic power of the leading order term. On the other hand
one also observes that the odd powers of $x$ have vanishing
coefficients. Note that for the cases I(a1) and I(a2) the
approximation of a massless electron is very precise.

The diagram classes I(b) and I(c) are constructed from the one-loop
diagram by adding either a two-loop and a one-loop fermion insertion
or a three-loop double-bubble diagram. We refrain from separating the
corresponding counterterm contribution for $\alpha$ and thus we
consider the sum of I(b) and I(c) which we denote by I(bc). In analogy
to case I(a) the case I(bc) is split into a contribution with two
electron loops [I(bc0)] and contributions with an electron and a muon
loop [I(bc1) and I(bc2)].  In case of I(bc1) the electron loop has
four photon couplings, in I(bc2) it is the muon loop.  Our results
read
\begin{align*}
A_2^\text{(8),I(bc0)} =\ & 0.558875 + 0.714610 \ell_x + 0.416667 \ell_x^2 \\
 & + x[-8.77188 \pm 0.00035] \\
 & + x^2[-51.3498 - 61.6586 \ell_x - 18 \ell_x^2 - 6.6667 \ell_x^3] \\
 & + x^3[176.8096 \pm 0.0014] \\
=\ & [0.55887 - 3.81001 + 11.84414] \\
 & + [-0.0424237 \pm 0.0000017] \\
 & + [-0.00120 + 0.00769 - 0.01197 + 0.02363] \\
 & + [0.00002000] \\
=\ & [8.59301] \\
 & + [-0.0424237 \pm 0.0000017] \\
 & + [0.018153] \\
 & + [0.00002000] \\
=\ & 8.568755 \pm 0.000002
\,,
\\
A_2^\text{(8),I(bc1)} =\ & 0.057516 \pm 0.001200 - 0.0156874 \ell_x \\
 & + x[0] \\
 & + x^2[0.30059 \pm 0.00017 + 0.359473 \ell_x] \\
 & + x^3[3.15081 - 0.292433 \ell_x] \\
=\ & [0.057516 \pm 0.001200 + 0.083639] \\
 & + [0] \\
 & + [0.0000070307 \pm 0.000000040 - 0.0000448286] \\
 & + [0.0000003564 + 0.0000001764] \\
=\ & [0.141155 \pm 0.001200] \\
 & + [0] \\
 & + [-0.0000377979 \pm 0.0000000040] \\
 & + [0.0000005328] \\
=\ & 0.1411 \pm 0.0012 
\,,
\\
A_2^\text{(8),I(bc2)} =\ & -0.068531 \pm 0.000075 - 0.105741 \ell_x \\
 & + x[0.0626838] \\
 & + x^2[0.422851 \pm 0.000053] \\
 & + x^3[-3.26780] \\
=\ & [-0.068531 \pm 0.000075 + 0.563770] \\
 & + [0.0003032] \\
 & + [0.0000098905 \pm 0.0000000012] \\
 & + [-0.0000003697] \\
=\ & [0.495239 \pm 0.000075] \\
 & + [0.0003032] \\
 & + [0.0000098905 \pm 0.0000000012] \\
 & + [-0.0000003697] \\
=\ & 0.495552 \pm 0.000075
\,.
\end{align*}
Similarly to the subsets of I(a) the diagram class I(bc0) is more than
one order of magnitude larger compared to I(bc1) and I(bc2). The
higher order corrections of I(bc0) is below one per cent and in case
of I(bc1) and I(bc2) even smaller. It is interesting to note that the
$x^2$ contribution of I(bc0) is around 40\% of the size of the $x^1$
part, though the order $x^3$ term is much smaller justifying the
truncation of the expansion.

Our result for diagram class I(d) reads
\begin{align*}
A_2^\text{(8),I(d)} =\ & -0.124375 + 0.03125 \ell_x \\
 & + x[17.4222 \pm 0.0997] \\
 & + x^2[40.1185 + 53.5293 \ell_x + 5.25 \ell_x^2 + 6 \ell_x^3] \\
 & + x^3[-349.570 \pm 2.148] \\
=\ & [-0.124375 - 0.166612] \\
 & + [0.08426 \pm 0.00048] \\
 & + [0.0009384 - 0.0066755 + 0.0034907 - 0.0212694] \\
 & + [-0.00003954 \pm 0.00000024] \\
=\ & [-0.290987] \\
 & + [0.08426 \pm 0.00048] \\
 & + [-0.0235159] \\
 & + [-0.00003954 \pm 0.00000024] \\
=\ & -0.23028 \pm 0.00048
\,.
\end{align*}
This contribution is an example where the first terms of the
asymptotic expansion provides a bad approximation to the final result.
In fact, the linear term is about 30\% of the leading order
contribution. Also the quadratic correction is
only about a factor four smaller than the linear term. However, the $x^3$ term is
at per mille level and justifies the assignment of a negligible
uncertainty from the asymptotic expansion.

The diagram classes II(a), II(b) and II(c) can be constructed from the
two-loop diagrams with two photons, where either a two-loop
electron self energy is inserted in one of the photon lines [II(a)],
a one-loop contribution is inserted in each photon line [II(b)] or 
two one-loop contribution are inserted in one of the photons [II(c)].
We combine again the contributions from II(b) and II(c) and denote by
II(bc0) and II(bc1) the cases with two electron loops and one electron
and one muon loop, respectively.
Our results are given by
\begin{align*}
A_2^\text{(8),II(a)} =\ & -0.934278 + 0.344166 \ell_x \\
 & + x[-2.83881] \\
 & + x^2[-9.44284 - 2.73820 \ell_x + 6 \ell_x^2] \\
 & + x^3[234.5798 \pm 0.0044 + 37.9774 \ell_x] \\
=\ & [-0.93428 - 1.83496] \\
 & + [-0.0137294] \\
 & + [-0.0002209 + 0.0003415 + 0.0039893] \\
 & + [0.0000265362 \pm 0.0000000005 - 0.0000229050] \\
=\ & [-2.76923] \\
 & + [-0.0137294] \\
 & + [0.004110] \\
 & + [0.0000036312 \pm 0.0000000005] \\
=\ & -2.77885
\,,
\\
A_2^\text{(8),II(bc0)} =\ & 0.307547 - 0.0974404 \ell_x - 0.458889 \ell_x^2 \\
 & + x[-2.97459 - 0.822467 \ell_x] \\
 & + x^2[8.84230 - 9.50786 \ell_x - 5.33333 \ell_x^2] \\
 & + x^3[-7.15473 \pm 0.00091 + 53.9477 \ell_x + 4.38649 \ell_x^2] \\
=\ & [0.3075 + 0.5195 - 13.0443] \\
 & + [-0.0143861 + 0.0212076] \\
 & + [0.000207 + 0.001186 - 0.003546] \\
 & + [-0.00000081 - 0.00003254 + 0.00001411] \\
=\ & [-12.2173] + [0.006822] + [-0.002154] + [-0.00001924] \\
=\ & -12.2126 
\,,
\\
A_2^\text{(8),II(bc1)} =\ & -0.0817145 \pm 0.0000026 + 0.300345 \ell_x \\
 & + x[-0.0239766] \\
 & + x^2[-0.0725930 \pm 0.0000015 + 0.125499 \ell_x] \\
 & + x^3[-0.323625] \\
=\ & [-0.081714 \pm 0.000003 - 1.601317] \\
 & + [-0.0001160] \\
 & + [-0.00000170 - 0.00001565 ] \\
 & + [-0.00000003661] \\
=\ & [-1.683031 \pm 0.000003] \\
 & + [-0.0001160] \\
 & + [-0.00001735] \\
 & + [-0.00000003661] \\
=\ & -1.683165 \pm 0.000003
\,.
\end{align*}
The numerically dominant class is II(bc0); the closed muon loop in
II(bc1) leads again to a result which is an order of magnitude
smaller. In both cases the corrections from higher orders in $x$ are
very small. The result of II(a) gets a contribution of nearly 0.5\%
from higher orders in $x$.

The asymptotic expansion of diagram class III shows a similar
behaviour: already the linear correction term is below 0.04\%
which is actually smaller than the numerical uncertainty from the
leading contribution. The result is given by
\begin{align*}
A_2^\text{(8),III} =\ & 1.15444 \pm 0.00446 - 1.80996 \ell_x \\
 & + x[-0.849197] \\
 & + x^2[-1.95556 \pm 0.00400 - 1.25333 \ell_x] \\
 & + x^3[-20.2365 - 15.3527 \ell_x] \\
=\ & [1.1544 \pm 0.0045 + 9.6500] \\
 & + [-0.004107] \\
 & + [-0.00004574 \pm 0.00000009 + 0.00015630] \\
 & + [-0.000002289 + 0.000009260] \\
=\ & [10.8044 \pm 0.0045] \\
 & + [-0.004107] \\
 & + [0.00011056 \pm 0.00000009] \\
 & + [0.000006970] \\
=\ & 10.8004 \pm 0.0045
\,.
\end{align*} 

Diagram class IV(d) is different from the previous cases since
it does not contain  photon self energy insertions. As a result
there is no logarithmic enhancement in the leading order term.
Note that IV(d) involves also more complicated master integrals
which lead to a larger relative uncertainty. The higher order
corrections in $x$ are again small. We obtain
\begin{align*}
A_2^\text{(8),{IV(d)}} =\ & -4.33491 \pm 0.06055 \\
 & + x[1.61430 \pm 0.09570] \\
 & + x^2[-435.46 \pm 13.85 + (-135.694 \pm 0.166) \ell_x] \\
 & + x^3[-328.90 \pm 24.77 - 614.619 \ell_x + 39.4784 \ell_x] \\
=\ & [-4.33491 \pm 0.06055] \\
 & + [0.007807 \pm 0.000463] \\
 & + [-0.010186 \pm 0.000324 + (0.016922 \pm 0.000021)] \\
 & + [-0.0000372 \pm 0.0000028 + 0.0003707 + 0.0001269] \\
=\ & [-4.335 \pm 0.061] \\
 & + [0.00781 \pm 0.00046] \\
 & + [0.00710 \pm 0.00037] \\
 & + [0.0004604 \pm 0.0000028] \\
=\ & -4.320 \pm 0.061
\,.
\end{align*}

Note that some diagram classes have been computed analytically in
the limit $m_e = 0$, with which our leading order results agree. In
Ref.~\cite{Lee:2013sx} the classes I(a0), I(a1) and the sum of I(bc0)
and II(bc0) are presented. The leading order of I(d) can be found in
Ref.~\cite{Broadhurst:1992za}.

\begin{table}[tb]
\begin{center}
\begin{tabular}{l@{\hskip -0.2cm}lll}
$A_2^{(8)}(m_\mu/m_e)$ & \hphantom{-00} this work & \hphantom{-00} literature & \\
\hline
I(a0) & $\hphantom{-00}7.223076$ & $\hphantom{-00}7.223077 \pm 0.000029$ & \cite{Kinoshita:2004wi} \\
 & & $\hphantom{-00}7.223076$ & \cite{Laporta:1993ds} \\
I(a1) & $\hphantom{-00}0.494072$ & $\hphantom{-00}0.494075 \pm 0.000006$ & \cite{Kinoshita:2004wi} \\
 & & $\hphantom{-00}0.494072$ & \cite{Laporta:1993ds} \\
I(a2) & $\hphantom{-00}0.027988$ & $\hphantom{-00}0.027988 \pm 0.000001$ & \cite{Kinoshita:2004wi} \\
 & & $\hphantom{-00}0.027988$ & \cite{Laporta:1993ds} \\
I(a) & $\hphantom{-00}7.745136$ & $\hphantom{-00}7.74547 \pm 0.00042$ & \cite{Aoyama:2012wk} \\
\hline
I(bc0) & $\hphantom{-00}8.56876 \pm 0.00001$ & $\hphantom{-00}8.56874 \pm 0.00005$  & \cite{Kinoshita:2004wi} \\
I(bc1) & $\hphantom{-00}0.1411 \pm 0.0060$ & $\hphantom{-00}0.141184 \pm 0.000003$  & \cite{Kinoshita:2004wi} \\
I(bc2) & $\hphantom{-00}0.4956 \pm 0.0004$ & $\hphantom{-00}0.49565 \pm 0.00001$  & \cite{Kinoshita:2004wi} \\
I(bc) & $\hphantom{-00}9.2054 \pm 0.0060$ & $\hphantom{-00}9.20632 \pm 0.00071$ & \cite{Aoyama:2012wk} \\
\hline
I(d) & $\hphantom{0}\text{$-$}\hphantom{0}0.2303 \pm 0.0024$ & $\hphantom{0}\text{$-$}\hphantom{0}0.22982 \pm 0.00037$ & \cite{Aoyama:2012wk} \\
 & & $\hphantom{0}\text{$-$}\hphantom{0}0.230362 \pm 0.000005$ & \cite{Baikov:1995ui} \\
\hline
II(a) & $\hphantom{0}\text{$-$}\hphantom{0}2.77885$ & $\hphantom{0}\text{$-$}\hphantom{0}2.77888 \pm 0.00038$ & \cite{Aoyama:2012wk} \\
 & & $\hphantom{0}\text{$-$}\hphantom{0}2.77885$ & \cite{Laporta:1993ds} \\
\hline
II(bc0) & $\hphantom{0}\text{$-$}12.212631$ & $\hphantom{0}\text{$-$}12.21247 \pm 0.00045$ & \cite{Kinoshita:2004wi} \\
II(bc1) & $\hphantom{0}\text{$-$}\hphantom{0}1.683165 \pm 0.000013$ & $\hphantom{0}\text{$-$}\hphantom{0}1.68319 \pm 0.00014$ & \cite{Kinoshita:2004wi} \\
II(bc) & $\hphantom{0}\text{$-$}13.895796 \pm 0.000013$ & $\hphantom{0}\text{$-$}13.89457 \pm 0.00088$ & \cite{Aoyama:2012wk} \\
\hline
III & $\hphantom{-0}10.800 \pm 0.022$ & $\hphantom{-0}10.7934 \pm 0.0027$ & \cite{Aoyama:2012wk} \\
\hline
IV(a0) & $\hphantom{-}116.76 \pm 0.02$ & $\hphantom{-}116.759183  \pm 0.000292$ & \cite{Kinoshita:2004wi} \\
 & & $\hphantom{-}111.1 \pm 8.1$ & \cite{Calmet:1975tw} \\
 & & $\hphantom{-}117.4 \pm 0.5$ & \cite{Chlouber:1977dr} \\
IV(a1) & $\hphantom{-00}2.69 \pm 0.14$ & $\hphantom{-00}2.697443 \pm 0.000142$ & \cite{Kinoshita:2004wi} \\
IV(a2) & $\hphantom{-00}4.33 \pm 0.17$ & $\hphantom{-00}4.328885 \pm 0.000293$ & \cite{Kinoshita:2004wi} \\
IV(a) & $\hphantom{-}123.78\pm 0.22$ & $\hphantom{-}123.78551 \pm 0.00044$ & \cite{Aoyama:2012wk} \\
\hline
IV(b) & $\hphantom{0}\text{$-$}\hphantom{0}0.38 \pm 0.08$ & $\hphantom{0}\text{$-$}\hphantom{0}0.4170 \pm 0.0037$ & \cite{Aoyama:2012wk} \\
IV(c) & $\hphantom{-00}2.94 \pm 0.30$  & $\hphantom{-00}2.9072 \pm 0.0044$ & \cite{Aoyama:2012wk} \\
\hline
IV(d) & $\hphantom{0}\text{$-$}\hphantom{0}4.32 \pm 0.30$ & $\hphantom{0}\text{$-$}\hphantom{0}4.43243 \pm 0.00058$ & \cite{Aoyama:2012wk} \\
\hline
\end{tabular}
\end{center}
\caption{Final results for the different classes and comparison
  with the literature. Note that the uncertainties in the second column
  are multiplied by a factor five.
  The results for IV(a)-IV(c) have been taken over from Ref.~\cite{Kurz:2015bia}.}
\label{tab::res}
\end{table}

Our final results are summarized in Table~\ref{tab::res}. As compared
to the expression in the above equations, which show the one standard
deviation uncertainties originating from the numerical integration, we
multiply the uncertainties in Table~\ref{tab::res} by a factor five, in
order to present conservative results. In case no uncertainty is
displayed the corresponding result is either known analytically or
with high numerical precision.

For comparison we show in the most right column of Table~\ref{tab::res} the
results available in the literature.  The most up-to-date results for the
diagram classes defined in Figure~\ref{fig::classes} can be found in
Ref.~\cite{Aoyama:2012wk}. To obtain separate results for sub-classes we
resort to Ref.~\cite{Kinoshita:2004wi}.  We could reproduce the analytic
results for some of the sub-classes which have been obtained in
Ref.~\cite{Laporta:1993ds}.  In addition we managed to obtain analytic results
for the case II(bc0). The corresponding results are shown in
Appendix~\ref{app::A}.

We find perfect agreement of our results with the ones in the
literature although in some cases the uncertainty is far
below the per mille level. In our approach we find relative 
large uncertainties of about {10\%} and {7\%} for the classes IV(c) and
IV(d), respectively. For class III the uncertainty amounts to {0.2\%} 
and for I(d) {1\%}.

Our final result for $A_2^{(8)}(m_\mu/m_e)$ is given by
\begin{eqnarray}
  A_2^{(8)} &=& A_2^{(8),\rm lbl} + A_2^{(8),\rm rem}
  \nonumber\\
  &=& 126.34(38) + 6.53(30) = 132.86(48)\,.
\end{eqnarray}
Our numerical uncertainty amounts to approximately ${0.5} \times
(\alpha/\pi)^4 \approx {1.5} \times 10^{-11}$. It is larger than
the uncertainty in Ref.~\cite{Aoyama:2012wk}.  Nevertheless it is
sufficiently accurate as can be seen by the comparison to the
difference between the experimental result and theory prediction which
is given by\footnote{This result is taken from
  Ref.~\cite{Aoyama:2012wk}.}
\begin{eqnarray}
  a_\mu({\rm exp}) - a_\mu({\rm SM}) &\approx& 249(87) \times 10^{-11}
  \,.
  \label{eq::amu_diff}
\end{eqnarray}
Note that the uncertainty in Eq.~(\ref{eq::amu_diff}) receives
approximately the same amount from experiment and theory (i.e. the
hadronic contribution). Even after a projected reduction of the
uncertainty by a factor four both in $a_\mu({\rm exp})$ and
$a_\mu({\rm SM})$ our numerical precision is a factor ten below
the uncertainty of the difference.


\section{\label{sec::res_e_tau}Simultaneous electron and tau contribution to $a_\mu$}

This section provides the four-loop results to $a_\mu$ from the Feynman diagrams
which contain simultaneously a closed electron and tau loop. Such contributions
arise from the diagrams classes I(a), I(b), I(c), II(b), II(c) and IV(a).

Each integral depends on all three lepton masses. As discussed in
Section~\ref{sec::calc} we perform a nested asymptotic expansion to
obtain a double expansion in $m_\mu/m_\tau$ and $m_e/m_\mu$.  Due to the
decoupling of the heavy tau lepton the whole contribution is suppressed
by a factor $1/m_\tau^2$. Furthermore, The expansion in the inverse
heavy tau mass is analytic and thus produces only even powers in
$m_\tau$. However, odd powers in $m_\mu/m_\tau$ and $m_e/m_\tau$ arise
from the two-loop mass counterterm
contribution~\cite{Gray:1990yh,Bekavac:2007tk}.  Thus, we obtain one
more order in $1/m_\tau$ for free which is taken into account in the
analytic and numeric results presented below.

In the following we present analytic results for
$A^{(8)}_{3}(m_\mu/m_e, m_\mu/m_\tau)$
including terms up to ${\cal O}(1/m_\tau^7)$.
We expand in $m_e/m_\mu$ up to the order where non-computed terms 
can safely be neglected. In fact, neglected terms are of order
$m_e^4 /m_\mu   /m_\tau^3$,
$m_e^3 m_\mu^2  /m_\tau^5$, and
$m_e^2 m_\mu^5  /m_\tau^7$.
Our result reads
\begin{align}\label{equ:eltau}
A^{(8)}_{3}\left(\frac{m_\mu}{m_e}, \frac{m_\mu}{m_\tau}\right) =\ &\frac{m_\mu^2}{m_\tau^2} \bigg(\frac{1}{135}\ln^2\frac{m_e^2}{m_\mu^2}+\frac{89}{810}\ln^2\frac{m_\mu^2}{m_\tau^2} +\ln\frac{m_\mu^2}{m_\tau^2} \left(\frac{22493}{291600}-\frac{3\zeta_3}{2}\right) \nonumber\\
 &\qquad+\ln\frac{m_e^2}{m_\mu^2} \left(-\frac{23}{270}\ln\frac{m_\mu^2}{m_\tau^2}-\frac{3 \zeta_3}{2}+\frac{2 \pi^2}{45}+\frac{74597}{97200}\right) \nonumber\\
 &\qquad+\frac{17 \zeta_3}{135} +\frac{2 \pi^4}{75}+\frac{193 \pi^2}{810}-\frac{984587}{486000}-\frac{8}{135} \pi^2\log(2)\bigg) \nonumber\\
 &+\frac{m_e m_\mu}{m_\tau^2} \left(\frac{4 \pi^2}{15}\ln\frac{m_\mu^2}{m_\tau^2}-\frac{821 \pi^2}{900}\right) \nonumber\\
 &+\frac{m_e^2}{m_\tau^2} \bigg(\ln^2\frac{m_e^2}{m_\mu^2} \left(-\frac{1}{10}\ln\frac{m_\mu^2}{m_\tau^2}-\frac{\pi^2}{36}+\frac{17}{50}\right) \nonumber\\
 &\qquad+\ln\frac{m_e^2}{m_\mu^2} \left(\frac{3}{5}\ln\frac{m_\mu^2}{m_\tau^2}-\frac{\zeta_3}{3}-\frac{4 \pi^4}{135}-\frac{23 \pi^2}{135}-\frac{673}{450}\right) \nonumber\\
 &\qquad+\left(-\frac{43}{45}-\frac{\pi^2}{15}\right) \ln\frac{m_\mu^2}{m_\tau^2} -\frac{16 \zeta_5}{9}-\frac{8 \pi^2 \zeta_3}{15}+\frac{47 \zeta_3}{45} \nonumber\\
 &\qquad-\frac{251 \pi^4}{4050}+\frac{56 \pi^2}{225}+\frac{45671}{8100}-\frac{4}{15} \pi^2 \log(2)\bigg) \nonumber\\
 &+\frac{m_e^3}{m_\mu m_\tau^2} \left(\frac{28 \pi^3}{135}-\frac{64 \pi^2}{135}\ln\frac{m_e^2}{m_\mu^2}-\frac{4 \pi^2}{135}\ln\frac{m_\mu^2}{m_\tau^2}-\frac{5689 \pi^2}{24300}\right) \nonumber\\
 &+\frac{\pi^2 }{90} \frac{m_e m_\mu^2}{m_\tau^3} \nonumber\\
 &+\frac{m_\mu^4}{m_\tau^4} \bigg(\ln^2\frac{m_e^2}{m_\mu^2} \left(\frac{1}{420}\ln\frac{m_\mu^2}{m_\tau^2}+\frac{3}{19600}\right) +\frac{1181}{40824}\ln^3\frac{m_\mu^2}{m_\tau^2}\nonumber\\
 &\qquad+\ln\frac{m_e^2}{m_\mu^2} \bigg(
\frac{4553}{90720}\ln^2\frac{m_\mu^2}{m_\tau^2}
-\frac{1074457}{4762800}\ln\frac{m_\mu^2}{m_\tau^2}-\frac{1811 \zeta_3}{2304} \nonumber\\
 &\qquad+\frac{1811 \pi^2}{68040}+\frac{2304926093}{2667168000}\bigg) -\frac{3034811}{38102400}\ln^2\frac{m_\mu^2}{m_\tau^2} \nonumber\\
 &\qquad+\ln\frac{m_\mu^2}{m_\tau^2} \left(-\frac{61849 \zeta_3}{80640}+\frac{3520386679}{4000752000}-\frac{2011 \pi^2}{204120}\right) \nonumber\\
 &\qquad-\frac{9564133 \zeta_3}{76204800}+\frac{50467 \pi^4}{2903040}+\frac{9308371 \pi^2}{85730400} \nonumber\\
 &\qquad-\frac{9932011422817}{5040947520000}-\frac{7}{1152}\log^4(2) \nonumber\\
 &\qquad+\frac{7 \pi^2}{1152}\log^2(2)-\frac{44}{945} \pi^2 \log(2)-\frac{7 a_4}{48}\bigg) \nonumber\\
 &+\frac{m_e m_\mu^3}{m_\tau^4} \left(\frac{19 \pi^2}{720}\ln\frac{m_\mu^2}{m_\tau^2}-\frac{2161 \pi^2}{43200}\right) \nonumber\\
 &+\frac{m_e^2 m_\mu^2}{m_\tau^4} \bigg(\ln^2\frac{m_e^2}{m_\mu^2} \left(\frac{61}{3528}-\frac{1}{84}\ln\frac{m_\mu^2}{m_\tau^2}\right) -\frac{3}{280}\ln^3\frac{m_\mu^2}{m_\tau^2}\nonumber\\
 &\qquad+\ln\frac{m_e^2}{m_\mu^2} \bigg(-\frac{9}{280}\ln^2\frac{m_\mu^2}{m_\tau^2}+\frac{49181}{176400}\ln\frac{m_\mu^2}{m_\tau^2}+\frac{\pi^2}{168} \nonumber\\
 &\qquad-\frac{5938843}{12348000}\bigg) +\frac{130813}{1058400}\ln^2\frac{m_\mu^2}{m_\tau^2} \nonumber\\
 &\qquad+\ln\frac{m_\mu^2}{m_\tau^2} \left(\frac{\zeta_3}{70}-\frac{1050211}{1543500}+\frac{17 \pi^2}{1260}\right) \nonumber\\
 &\qquad+\frac{6751 \zeta_3}{14700}+\frac{\pi^4}{1050}-\frac{163823 \pi^2}{1587600}-\frac{271714897}{23337720000}\bigg) \nonumber\\
 &+\frac{m_e m_\mu^4}{m_\tau^5} \left(\frac{\pi^2}{140}\ln\frac{m_\mu^2}{m_\tau^2}+\frac{79 \pi^2}{19600}\right) \nonumber\\
 &+\frac{m_\mu^6}{m_\tau^6} \bigg(\ln^2\frac{m_e^2}{m_\mu^2} \left(\frac{2}{945}\ln\frac{m_\mu^2}{m_\tau^2}-\frac{131}{297675}\right) \nonumber\\
 &\qquad+\ln\frac{m_e^2}{m_\mu^2}\bigg(\frac{20929}{604800}\ln^2\frac{m_\mu^2}{m_\tau^2}-\frac{229507}{1984500}\ln\frac{m_\mu^2}{m_\tau^2} \nonumber\\
 &\qquad-\frac{3077 \zeta_3}{5760}+\frac{83 \pi^2}{4800}+\frac{84725571607}{160030080000}\bigg) \nonumber\\
 &\qquad+\frac{15787}{777600}\ln^3\frac{m_\mu^2}{m_\tau^2}-\frac{104754659}{2286144000}\ln^2\frac{m_\mu^2}{m_\tau^2} \nonumber\\
 &\qquad+\ln\frac{m_\mu^2}{m_\tau^2} \left(-\frac{62761 \zeta_3}{120960}+\frac{33670638521}{51438240000}-\frac{3973 \pi^2}{544320}\right) \nonumber\\
 &\qquad-\frac{34589999 \zeta_3}{152409600}+\frac{818557 \pi^4}{65318400}+\frac{43896581 \pi^2}{685843200} \nonumber\\
 &\qquad-\frac{34820138253959}{28355329800000}-\frac{17}{2880}\log^4(2)+\frac{17 \pi^2}{2880}\log ^2(2) \nonumber\\
 &\qquad-\frac{512 \pi^2}{14175}\log(2) -\frac{17 a_4}{120}\bigg) \nonumber\\
 &+\frac{m_e m_\mu^5}{m_\tau^6} \left(\frac{11 \pi^2}{5670}\ln\frac{m_\mu^2}{m_\tau^2}-\frac{39157 \pi^2}{3572100}\right) \nonumber\\
 &+\frac{m_e m_\mu^6}{m_\tau^7} \left(\frac{\pi^2}{105}\ln\frac{m_\mu^2}{m_\tau^2}+\frac{79 \pi^2}{66150}\right)
 + \ldots\,.
\end{align}

For the numerical evaluation of our result we use~\cite{Beringer:1900zz}
\begin{eqnarray}
  m_\mu/m_e &=& 206.7682843(52)\,,
  \nonumber\\
  m_\mu/m_\tau &=& 5.94649(54) \cdot 10^{-2}
  \,,
\end{eqnarray}
and obtain
\begin{align}
  A^{(8)}_{3,\mu}(m_\mu/m_e, m_\mu/m_\tau) &\approx 0.06321803 -
  0.00049494 - 0.00000111 \nonumber\\ &= 0.0627220(1)(100)
  \,,
  \label{eq::A8emt}
\end{align}
where the three numbers in the first line are obtained as follows:
The first number includes terms up to ${\cal O}(m_\tau^{-3})$
with numerator factors $m_e^0$ or $m_e^1$. All quadratic corrections in
$m_e$ are part of the second term which also contains all ${\cal
  O}(m_\tau^{-4})$ and ${\cal O}(m_\tau^{-5})$ corrections. The last term in
the first line of~(\ref{eq::A8emt}) comprises the remaining contributions of
order ${\cal O}(m_\tau^{-6})$ and ${\cal O}(m_\tau^{-7})$ and the cubic
electron mass terms.
In the second line the final result is presented. The first uncertainty is set
to 10\% of the last expansion term in the first line which is our estimate for
the missing higher order terms.  The second error originates from the
experimental uncertainties for $m_\mu/m_\tau$ and $m_\mu/m_e$ as given above.
The result is in good agreement with $A^{(8)}_{3}(m_\mu/m_e, m_\mu/m_\tau) =
0.06272(4)$~\cite{Aoyama:2012wk}.

\begin{table}[tb]
  \begin{center}
    \begin{tabular}{cll}
      & \multicolumn{2}{c}{$A^{(8)}_{3}(m_\mu/m_e, m_\mu/m_\tau)$}\\
      \hline
      group & \hphantom{$-$}this work & \hphantom{$-$}Ref.~\cite{Aoyama:2012wk} \\
      \hline
      I(a) & \hphantom{$-$}0.00320905(1) & \hphantom{$-$}0.003209(0) \\
      I(b) + I(c) & \hphantom{$-$}0.00442289(2) & \hphantom{$-$}0.004422(0)
      \\
      II(b) + II(c) & $-0.02865753(1)$ & $-0.028650(2)$ \\
      IV(a) & \hphantom{$-$}0.08374757(9) & \hphantom{$-$}0.083739(36)
    \end{tabular}
  \end{center}
  \caption{\label{tab::e_tau}Comparison of our results to Ref.~\cite{Aoyama:2012wk}
    for the individual diagram classes. Note that the error
      in $m_\mu/m_\tau$ is not included. In most cases it induces an
      uncertainty in fifth significant digit in the displayed numbers,
      in the cass of IV(a) even in the fourth. In the second row an updated
      result for I(c) has been used, see footnote~\ref{fnt:4}.}
\end{table}

Although the absolute contribution from $A^{(8)}_{3}$ is quite
small it is nevertheless instructive to compare our results to the
complete numerical method of Ref.~\cite{Aoyama:2012wk}.  This is
done in Table~\ref{tab::e_tau} for the individual diagram classes
contributing to $A^{(8)}_{3}(m_\mu/m_e, m_\mu/m_\tau)$.\footnote{We thank
  Makiko Nio for providing us with an updated result for the 
  contribution I(c).\label{fnt:4}}  Good agreement within the uncertainties are
found for the diagram classes I(a) and IV(a). For II(b)+II(c) one
observes a discrepancy of about three standard deviations. Also
the results for I(b)+I(c) do not agree within the assigned
uncertainty. Note, however, that in~\cite{Aoyama:2012wk} an older
value for $m_\mu/m_\tau$ has been used which is about 0.01\% smaller and
thus can explain most of the discrepancy.


\section{\label{sec::concl}Conclusions}

We presented results for the complete four-loop electron contribution
to the anomalous magnetic moment of the muon. This includes
light-by-light-type contributions (which have already been presented
in Ref.~\cite{Kurz:2015bia}), the remaining contributions where the
external photon couples to the (external) muon line (see
Section~\ref{sec::res}), and the contribution which involves both
electron and tau loops (see Section~\ref{sec::res_e_tau}).

Our calculation serves as an important cross check to the calculation
performed in Ref.~\cite{Aoyama:2012wk} which is based on an entire
numerical method. In our approach we express $a_\mu$ in terms of an
analytic linear combination of four-loop master integrals. Thus, our
result can be improved systematically by evaluating more and more
master integrals either analytically or with higher numerical precision.

We have demonstrated that our current (numerical) result for $a_\mu$
is sufficiently precise: even after reducing the current experimental
and theory uncertainty by a factor four the uncertainty induced by our
four-loop expression is a factor ten smaller.

As a by-product we have computed three-loop corrections to the mass
and wave function on-shell renormalization constants 
of the muon taking into account effects from a finite electron mass.

Note that the four-loop contribution involving heavy tau loops have been
computed in Ref.~\cite{Kurz:2013exa}. Thus, only the four-loop
universal part remains to be cross checked.

The four-loop results of Refs.~\cite{Kurz:2013exa,Kurz:2015bia}
and this paper can be summarized as
\begin{eqnarray}
  a_\mu^{(8)} &=& a_\mu^{(8)}|_{\rm univ.} 
  + 132.86(48) + 0.0424941(53) + 0.062722(10)
  \,,
\end{eqnarray}
where the second third and fourth term on the right-hand side
corresponds to $A_2^{(8)}(m_\mu/m_e)$, $A_2^{(8)}(m_\mu/m_\tau)$ and
$A_3^{(8)}(m_\mu/m_e,m_\mu/m_\tau)$, respectively.



\section*{Acknowledgments}

We would like to thank M.~Nio for useful communications concerning
the results from Ref.~\cite{Aoyama:2012wk}.
We thank the High Performance Computing Center
Stuttgart (HLRS) and the Supercomputing Center of Lomonosov Moscow
State University~\cite{LMSU} for providing computing time used for the
numerical computations with {\tt FIESTA}.  P.M was supported in part
by the EU Network HIGGSTOOLS PITN-GA-2012-316704. 
The work of V.S. was supported by the Alexander von
Humboldt Foundation (Humboldt Forschungspreis).


\appendix


\section{\label{app::A} Analytic results for selected contributions}

In this appendix we present results for the 
diagram classes I(a), II(a) and II(bc). The expansion coefficients are
either known analytically or to high numerical precision. They are
given by
\begin{align*}
A_2^\text{(8),I(a0)} =\,
 &-\frac{8609}{5832} - \frac{25 \pi^2}{162} - \frac{2 \zeta_3}{9}  + \left(-\frac{317}{162} - \frac{2 \pi^2}{27}\right) \ell_x - \frac{25}{27}\,\ell_x^2 - \frac{4}{27}\,\ell_x^3\\
 &+ 6.405155074501456\ x\\
 &+ x^2 \left(\frac{967}{315} + \frac{26 \pi^2}{27} + \frac{136 \zeta_3}{35} + \left(\frac{304}{27} + \frac{8 \pi^2}{9}\right) \ell_x + \frac{52}{9} \ell_x^2 + \frac{16}{9} \ell_x^3 \right)\\
 &-52.00224911971472\ x^3\\
A_2^\text{(8),I(a1)} =\,
 &\frac{7627}{1944} + \frac{13 \pi^2}{27} - \frac{4 \pi^4}{45} + \ell_x \left(\frac{61}{81} - \frac{2 \pi^2}{27}\right) + \ell_x^2 \left(\frac{119}{27} - \frac{4 \pi^2}{9}\right)\\
 &+x^2 \left(\frac{227}{18} - \frac{4 \pi^2}{3} + \ell_x \left(\frac{230}{27} - \frac{8 \pi^2}{9}\right)\right)\\
A_2^\text{(8),I(a2)} =\,
 &0.0007061505929186751 + \ell_x \left(\frac{943}{162} + \frac{8 \pi^2}{135} - \frac{16 \zeta_3}{3}\right)\\
 &+ 0.004933870387556993\ x^2\\
A_2^\text{(8),II(a)} =\,
 &-0.9342776853294055 + \ell_x \left(\frac{31}{16} - \frac{5 \pi^2}{12} + \frac{\pi^2 \log(2)}{2} - \frac{3 \zeta_3}{4}\right)\\
 &+ x \left(\frac{79 \pi^2}{54} - \frac{13 \pi^3}{36} - \frac{8 \pi^2 \log(2)}{9}\right)\\
 &+ x^2 \bigg(-9.442838330353652\\
 &\qquad+ \ell_x \left(-\frac{47}{2} + \frac{37 \pi^2}{4} - 14 \pi^2 \log(2) + 21 \zeta_3\right) + 6\,\ell_x^2 \bigg)\\
A_2^\text{(8),II(bc0)} =\,
&\frac{2299 \pi ^2}{648}-\frac{17233}{1728}+\frac{40 a_4}{3}+\frac{16 a_5}{3}-\frac{37 \zeta_5}{6}-\frac{\pi ^2 \zeta_3}{8}+\frac{431 \zeta_3}{36}\\
&-\frac{403 \pi^4}{3240}-\frac{2 \log^5(2)}{45}+\frac{5 \log ^4(2)}{9}-\frac{4}{27} \pi ^2 \log^3(2)\\
&-\frac{31}{540} \pi ^4 \log (2)-\frac{235}{54} \pi ^2 \log (2)+\frac{10}{9} \pi ^2 \log ^2(2) \\
&+\bigg(\frac{16 a_4}{3}+7 \zeta_3-\frac{11 \pi ^4}{108}+\frac{79 \pi ^2}{27}-\frac{115}{12}\\
&\qquad+\frac{2 \log ^4(2)}{9}+\frac{4}{9} \pi ^2 \log^2(2)-\frac{10}{3} \pi ^2 \log (2)\bigg)\,\ell_x \\
&+\left(\zeta_3-\frac{31}{12}+\frac{5 \pi^2}{9}-\frac{2}{3} \pi ^2 \log (2)\right)\,\ell_x^2 +x \left(-\frac{217 \pi ^2}{720}-\frac{\pi ^2}{12}\,\ell_x\right) \\
&+x^2 \bigg(\frac{425}{36}-\frac{112 a_4}{3}-\frac{455 \zeta_3}{12}+\frac{335 \pi ^4}{432}-\frac{409 \pi ^2}{36}\\
&\qquad\quad-\frac{14 \log ^4(2)}{9}-\frac{28}{9} \pi ^2 \log ^2(2)+\frac{299}{18} \pi ^2 \log (2) \\
&\qquad\quad+\left(\frac{13}{3}-14 \zeta_3-\frac{37 \pi ^2}{6}+\frac{28}{3} \pi ^2 \log (2)\right)\,\ell_x-\frac{16}{3}\,\ell_x^2\bigg)
\end{align*}
Note that analytic results for I(a0), I(a1), I(a2) and II(a) can be
found in Ref.~\cite{Laporta:1993ds}.




\end{document}